# Anisotropic gap structure and sign reversal symmetry in monolayer Fe(Se,Te)

Yu Li,[†] Dingyu Shen,[†] Andreas Kreisel,[‡] Cheng Chen,[†] Tianheng Wei,[†] Xiaotong Xu[†] and Jian Wang[†,§,‖,#,*]

[†]International Center for Quantum Materials, School of Physics, Peking University, Beijing 100871, China.

[‡]Institut für Theoretische Physik, Universität Leipzig, D-04103 Leipzig, Germany.

[§]Collaborative Innovation Center of Quantum Matter, Beijing 100871, China.

[‖]CAS Center for Excellence in Topological Quantum Computation, University of Chinese Academy of Sciences, Beijing 100190, China.

[#]Beijing Academy of Quantum Information Sciences, Beijing 100193, China.

*Email: jianwangphysics@pku.edu.cn.

**ABSTRACT:** The iron-based superconductors are an ideal platform to reveal the enigma of the unconventional superconductivity and potential topological superconductivity. Among them, the monolayer Fe(Se,Te)/SrTiO$_3$(001), which is proposed to be topological nontrivial, shows interface-enhanced high-temperature superconductivity in the two dimensional limit. However, the experimental studies on the superconducting pairing mechanism of monolayer Fe(Se,Te) films are still limited. Here, by measuring quasiparticle interference in monolayer Fe(Se,Te)/SrTiO$_3$(001), we report the observation of the anisotropic structure of the large superconducting gap and the sign change of the superconducting gap on different electron pockets. The results are well consistent with the 'bonding-antibonding' $s_\pm$-wave pairing symmetry driven by spin fluctuations in conjunction with spin-orbit coupling. Our work is of basic significance not only for a unified superconducting formalism in the iron-based superconductors, but also for understanding of topological superconductivity in high-temperature superconductors.

**KEYWORDS:** *high-temperature superconductors, anisotropic superconducting gap, $s_\pm$-wave pairing symmetry, quasiparticle interference, scanning tunneling spectroscopy*

Among iron-based superconductors (FeBSs), the monolayer FeSe film grown on SrTiO$_3$(001) substrate (1-UC FeSe/STO) with an impressively large superconducting (SC) gap of almost 20 meV[1] and onset SC transition temperature ($T_c$) of above 40 K[2], has fueled intensive researches upon its SC properties and pairing mechanism[3-8]. Due to the electron doping, the hole pocket in the center (Γ) of the Brillouin zone (BZ) sinks below the Fermi energy in 1-UC FeSe/STO[9], which challenges the widely accepted $s_\pm$-wave pairing within the electron-hole Fermi pocket nesting picture in FeBSs[10]. Alternative pairing symmetries including $s_{++}$-wave[11,12], 'bonding-antibonding' $s_\pm$-wave[13,14], and 'quasi-nodeless' $d$-wave[15,16], have been theoretically proposed, but a consensus of the pairing symmetry is still lacking[17-21]. Recently, 1-UC Fe(Se,Te)/STO has been predicted to be topological nontrivial with $T_c$ comparable to that of the 1-UC FeSe/STO and much higher than bulk Fe(Se,Te)[22-28]. However, the researches on the pairing mechanism in 1-UC Fe(Se,Te)/STO are very limited. The experimental results related to gap structures and a direct visualization of gap signs, which are prerequisites to reveal the pairing mechanism, are highly desired for understanding both high $T_c$ and topological superconductivity in 1-UC Fe(Se,Te)/STO.

Quasiparticle interference (QPI) is designed for measuring the modulation of the differential conductance mappings $g(\boldsymbol{r},E) = dI/dV(\boldsymbol{r},eV)$, resulting from interference of quasiparticles scattered by defects. After the Fourier transform (FT), the resultant FT-QPI signal is a complex function expressed as $g(\boldsymbol{q},E) = |g(\boldsymbol{q},E)|\exp{(i\theta_{\boldsymbol{q},E})}$, where $\boldsymbol{q}$ is the momentum transfer of the scattering momenta $\boldsymbol{k}$ on the Fermi pockets ($\boldsymbol{q} = \boldsymbol{k}_1 - \boldsymbol{k}_2$). The intensity of the FT-QPI patterns $|g(\boldsymbol{q},E)|$ can provide information about the shape of the Fermi surfaces (FSs) and the momentum space ($\boldsymbol{k}$-space) structure of SC gaps[29,30]. Moreover, the scattering phase $\theta_{\boldsymbol{q},E}$ encodes messages about the sign of SC gaps, which can be analyzed by phase-referenced QPI (PR-QPI) methods[31-33]. The sign of SC gaps is critical to distinguish the gap symmetry and further deduce the pairing interaction. Therefore, QPI measurement is a powerful tool to study the amplitude and symmetry of SC order parameters and can reveal the SC pairing mechanism[34,35].

In this work, we measured and analyzed the FT-QPI patterns in 1-UC Fe(Se,Te)/STO to investigate the SC pairing mechanism. By fitting the tunneling spectra and utilizing the FT-QPI patterns to quantitatively extract the gap structure, we detected a highly anisotropic SC gap on the outer electron pocket at the corner ($\widetilde{M}$) of BZ. We further proposed a PR-QPI method and used it to verify the sign change between the SC gaps on the outer and inner electron pockets. The determination of the anisotropic gap and visualization of the sign reversal gaps are well consistent with 'bonding-antibonding' $s_\pm$-wave scenario[13,14]. Our work provides a valuable knowledge for a high $T_c$ topological superconductor candidate and contributes to a unified understanding of pairing mechanism in FeBSs.

The 1-UC Fe(Se,Te) film was epitaxially grown on a SrTiO$_3$(001) substrate by molecular beam epitaxy[26] (Supporting Information I.A). The nominal ratio of Se/Te (0.5/0.5) was estimated from the thickness of the second-layer Fe(Se,Te) film[25,26,36] (Supporting Information II.A). The scanning tunneling microscopy (STM) topographic image shows the topmost Se/Te atom arrangement of the 1-UC Fe(Se,Te) film (Figure 1a). The square lattice of top layer Se/Te atom corresponds to the two-iron (2-Fe) unit cell, which is depicted as the dashed black box in the crystal structure as shown in Figure 1b. A typical tunneling spectrum after normalization and symmetrization exhibits two pairs of coherence peaks and remarkable U-shaped SC gaps at 4.3 K, shown as the blue hollow dots in Figure 1c. The two-band Dynes model with anisotropic SC gap functions expresses reasonable fittings (red curve in Figure 1c) with pairing strengths of $|\Delta_1^{max}|$ = 9.55 ± 0.15 meV and $|\Delta_2^{max}|$ = 18.13 ± 0.08 meV, which are consistent with the energies of $|\Delta_1|$ = 11.0 meV and $|\Delta_2|$ = 18.4 meV determined by coherence peaks (also see Figure S1c, Supporting Information I.B). As for two-band Dynes model fitting without incorporating SC gap anisotropy[37,38], while the tunneling spectrum can be fitted, the obtained parameters of $|\Delta_1^{con}|$ = 7.87 ± 0.04 meV and $|\Delta_2^{con}|$ = 16.64 ± 0.08 meV are too small and physically irrational (see different types of Dynes model in Supporting Information II.B). Furthermore, we averaged over 180 spectra to represent a general tunneling spectrum, shown as the blue hollow dots in Figure 1d. Due to the average broadening, the SC gap of $|\Delta_1|$ shows kink structure rather than coherence peak. We repeated the Dynes model fittings for the averaged tunneling spectrum (red curve in Figure 1d), which yields almost the same anisotropic gap structure, as shown in the insets of Figure 1c and Figure 1d. This control experiment validates the universality of the gap anisotropy.

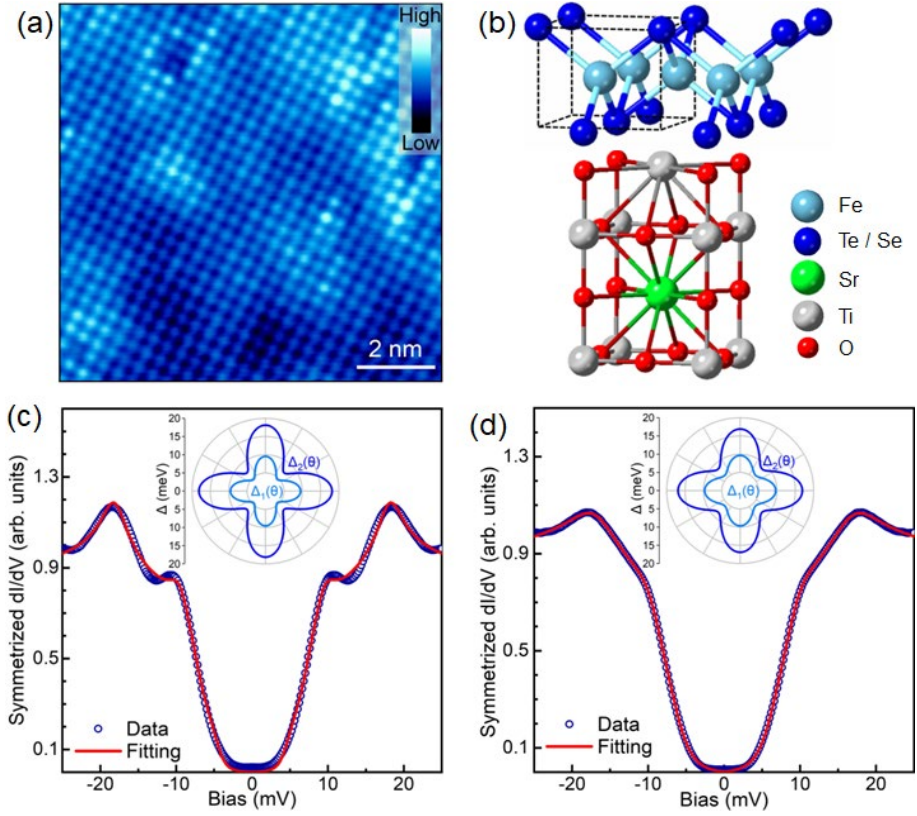

**Figure 1.** STM topography and SC gap anisotropy revealed by Dynes model fittings in 1-UC Fe(Se,Te)/STO. (a) STM topographic image (10×10 $nm^2$) showing the topmost Se/Te lattice. (b) Crystal structure of 1-UC Fe(Se,Te)/STO. The dashed black box indicates 2-Fe unit cell. (c) A typical d$I$/d$V$ spectrum after normalization and symmetrization (blue hollow dots) at 4.3 K. The red solid line is the theoretical fitting by two-band anisotropic Dynes model. Inset shows the anisotropic gap functions used in the fitting. (d) An averaged (over 180 spectra) tunneling spectrum with same representation as (c). The energies of $|\Delta_1|$ = 12.4 meV and $|\Delta_2|$ = 18.0 meV (determined by $d^2I/dV^2$ spectrum in Figure S1d, Supporting Information I.B) could represent the general SC gap value in the 1-UC Fe(Se,Te) film.

The Bogoliubov quasiparticles, elastically scattered by defects (see defect information in Supporting Information II.A), form the standing wave patterns which can be visualized by differential conductance mappings $g(r,E)$ (Figure 2a). The obtained FT-QPI pattern $|g(q,E)|$ (see the QPI measurements in Supporting Information III) shows three types of ring-like scattering structures (Figure 2c), which can be understood by the FSs in the folded BZ (Figure 2b). We discuss the $q_2$ and $q_3$ rings (Figure 2c), which are contributed by scattering vectors between electron pockets at the nearest-neighboring and next nearest-neighboring $\widetilde{M}$ point respectively, in Supporting Information IV.A. In the following, we concentrate on $q_1$ ring (Figure 2d), associated with scattering vectors within electron pockets at the same $\widetilde{M}$ point (Supporting Information IV.B). Due to the band hybridization deriving from spin-orbit coupling (SOC)[11,14,16], the two elliptic-shaped electron pockets in the folded BZ shall be recombined into inner and outer pockets marked as $\delta_1$ and $\delta_2$ in Figure 2b, respectively. Accordingly, $q_1$ ring should have three sub-sets, including intra-pocket scatterings of inner and outer pockets ($\delta_1 \leftrightarrow \delta_1$, $\delta_2 \leftrightarrow \delta_2$) as well as inter-pocket scatterings between them ($\delta_1 \leftrightarrow \delta_2$). Although the three sub-sets of $q_1$ ring are not distinctively resolved in our FT-QPI patterns, by extracting the line cuts of $|g(q,E)|$ along different radial direction $\theta$ and locating the maximum of each line cut (Figure 2e), we can determine the contour of $q_1$ scattering pattern (blue contour in Figure 2d) and obtain the FT-QPI intensity $I(\theta, E)$ along the contour. Essentially, the $I(\theta, E)$ is a projection of the $|g(q, E)|$ onto the band dispersion $E(q) = E(2k)$, concurrently mapping the coordinate from two-dimensional $q$ to one-dimensional $\theta$. We further implemented two steps to translate the raw intensity $I(\theta, E)$ into a normalized intensity $I_n(\theta, E)$ (Figure 2f), which unifies the FT-QPI intensity measured at different field of views and simultaneously preserves the energy dependence of FT-QPI intensity (Supporting Information V.A).

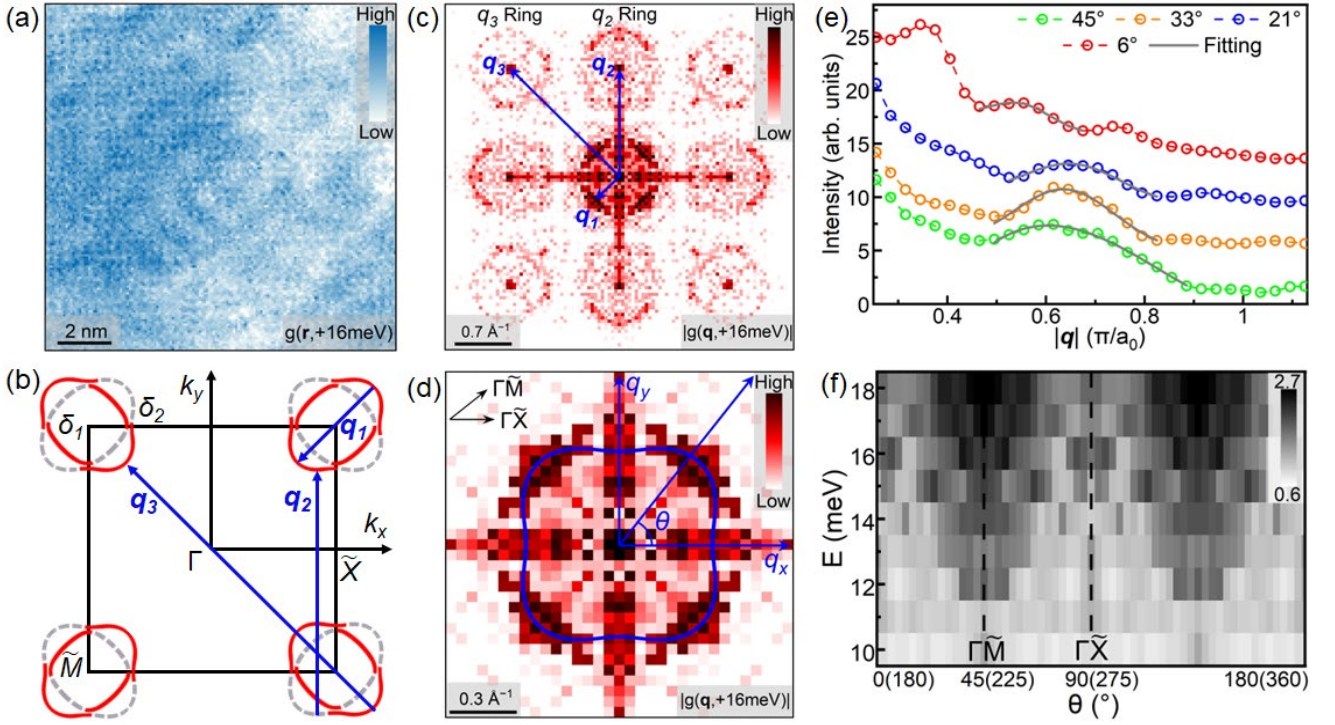

**Figure 2.** QPI and FT-QPI patterns. (a) Differential conductance mapping $g(r, E = 16 \text{ meV})$ (4.3 K, 13×13 nm$^2$) after distortion-correction, showing the interference patterns. (b) The schematic FSs of 1-UC Fe(Se,Te)/STO in the folded BZ. The red solid ellipses and gray dashed ellipses both denote the electron pockets, while the latter represents the folded bands. After band hybridization, the initial elliptic-shaped electron pockets form the inner pockets $\delta_1$ and outer pockets $\delta_2$. The blue solid lines $q_1$, $q_2$ and $q_3$ show the representative scattering vectors corresponding to the scattering vectors in (c). (c) FT-QPI pattern of $|g(q, E = 16 \text{ meV})|$ after symmetrization and Gaussian core subtraction, shows three types of scattering patterns. (d) Zoom-in image of the central part of (c) to clearly show $q_1$ scattering pattern. The blue closed curve [0.624 - 0.065 * cos (4$\theta$) $\pi/a_0$] depicts the contour of $q_1$ scattering pattern. (e) The line cuts of the raw FT-QPI intensity in (d) versus $q$ along different radial direction $\theta$ = 6°, 21°, 33° and 45° (vertically offset for clarity). The gray solid lines are the Gaussian fits plus linear background. The obtained positions of the Gaussian peak are 0.552 $\pi/a_0$ (6°), 0.667 $\pi/a_0$ (21°), 0.652 $\pi/a_0$ (33°) and 0.668 $\pi/a_0$ (45°). The intensities $I(\theta, E)$ of the Gaussian peak are further processed to give $I_n(\theta, E)$. (f) Two-dimensional map of $I_n(\theta, E)$. The FT-QPI intensity in the $\Gamma\tilde{X}$ directions is weaker than that of $\Gamma\tilde{M}$ directions.

When the SC gap ($\Delta(k)$) is anisotropic in $k$-space, the constant-energy contours (CECs) at the energy ($E$) within gap minimum and gap maximum, are roughly banana-shaped (red contours in Figure 3a), and the tips of the CECs, i.e. 'banana' tips (red solid dots in Figure 3a) possess a gap value equal to the energy of CECs ($\Delta(k) = E$)[29,30]. More importantly, since the locus of the 'banana' tips have the highest spectral weight, the scattering vectors connecting the tips should have maximum intensity in FT-QPI pattern[29,30]. Then, by tracking the evolution of the maximum-intensity points associated with these 'banana' tips, one can quantitatively extract the $k$-space structure of the anisotropic gap and simultaneously obtain the shape of the FS[29,30]. Herein, we adopted an approach[29], which tracks these maximum-intensity points along the gap minimum direction (also a high-symmetry direction). At an energy $E$ within anisotropic gap value, the scattering vectors naturally have two different lengths and produce two maximum-intensity points (shorter arrows $q_M^{t_1}$ and longer arrows $q_M^{t_2}$ with same color in Figure 3a,b). When the $E$ changes from gap minimum ($|\Delta_2^{min}|$) to gap maximum ($|\Delta_2^{max}|$), these maximum-intensity points lie on a curve (red dashed curve in Figure 3b), which continues with the band dispersion $E(q) = E(2k)$ (orange dashed curve in Figure 3b). The red curve can reflect the anisotropic structure of the SC gap and the FS (see the method to obtain FS in Supporting Information V.B), whereas the orange curve is not associated with the 'banana' tips but rather derives from the opposite-side scattering ($k \leftrightarrow -k$) between the CECs at the gap minimum direction.

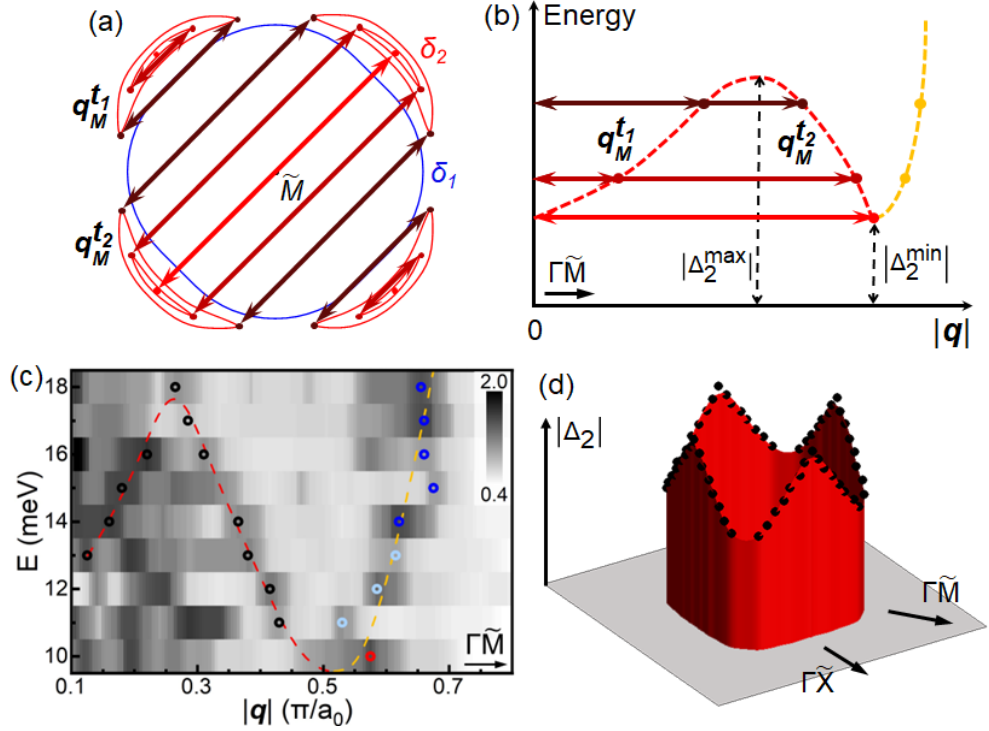

**Figure 3.** Anisotropic SC gap extracted from central FT-QPI pattern. (a) Schematic CECs of Bogoliubov quasiparticles (red contours) for $\delta_2$ at $\widetilde{M}$ in the folded BZ. The anisotropic gap is C$_4$ symmetric with maximum in the $\Gamma\widetilde{X}$ directions ($\theta = 0° + 90° \times N$) and minimum in the $\Gamma\widetilde{M}$ directions ($\theta = 45° + 90° \times N$). As energy increases, the size of CECs enlarges, and the tips of CECs are emphasized by the red dots with gradually darkened color. Correspondingly, the arrows with gradually darkened color depict the scatterings vectors associated with 'banana' tips along a high symmetry direction at $\theta = 45°$ ($\Gamma\widetilde{M}$). The blue contour represents the schematic FS of $\delta_1$. (b) The schematic curved shape in a $|q|$-$E$ plot. The red dashed curve, coming from the scattering vectors in (a) (shorter arrows $q_M^{t_1}$ and longer arrows $q_M^{t_2}$), reflects the anisotropic gap structure. The orange dashed curve shows the band dispersion $E(q) = E(2k)$. (c) Intensity plot in a $|q|$-$E$ plane where line cuts are along $\theta = 45°$. The black and colored hollow dots locate the maximum-intensity points associated with 'banana' tips and opposite-side scatterings $k \leftrightarrow -k$, respectively. The color of the dots indicates the sign in their corresponding PR-QPI pattern (Figure 4) with blue denoting negative and red denoting positive. Here, the light blue also denotes negative signals, albeit close to zero. The red and orange dashed curves schematically depict the trajectories of the maximum-intensity points, qualitatively matching the curves in (b). (d) The $k$-space structure of the SC gap (black solid dots) and FS, simultaneously obtained by (c).

Since the energy value of the SC coherence peak in the tunneling spectrum in principle corresponds to the gap maximum[29,30], we can recognize that the gap maximum is $|\Delta_2| = 18.0$ meV from the average tunneling spectrum (Figure 1d). When energy below the gap maximum, FT-QPI intensity is supposed to be suppressed along the gap maximum direction[29,39]. By extracting the intensity $I_n(\theta, E)$ of $q_1$ scattering pattern, we observed that the FT-QPI intensity in the $\Gamma\tilde{X}$ directions ($\theta = 0° + 90° \times N$) is weaker than that of $\Gamma\tilde{M}$ directions (Figure 2f). Therefore, we extracted the line cuts of FT-QPI intensity along the radial direction at $\theta = 45°$ ($\Gamma\tilde{M}$), i.e. the gap minimum direction. The intensity is plotted in a $|q|$-$E$ plane (Figure 3c). By locating the maximum-intensity points in the $|q|$-$E$ plane (hollow dots in Figure 3c), we obtained the two curved trajectories (red and orange dashed curves in Figure 3c) as expected. The orange dashed curve in Figure 3c shows the electron-like branch of the Bogoliubov spectrum. The red dashed curve in Figure 3c connects maximum-intensity points associated with 'banana' tips, which contains the information about the gap structure. Firstly, the peak of the trajectory reveals that the gap maximum is around 18 meV, which is consistent with the energy of the coherence peak in the averaged tunneling spectrum (Figure 1d). Secondly, the gap minimum determined by the intersection point of the two trajectories is around 10 meV, which can be verified by the PR-QPI patterns as shown in the following section. Thirdly, we show the measured $k$-space structure of the anisotropic SC gap on the simultaneously obtained FS in Figure 3d. The FS is square-shape, in accordance with the theoretical geometry of the outer pocket $\boldsymbol{\delta_2}$ formed by SOC induced band hybridization[11,14,16]. As discussed in Supporting Information V.C, another approximate approach[40] is adopted to double check the gap structure. A comparison of Figure S11 in Supporting Information V.C with Figure 3 reveals that both methods yield the same result.

One may argue that the maximum-intensity points in Figure 3c also appear at $E$ from 10 meV to 12 meV and within $|q|$ from 0.2 $\pi/a_0$ to 0.3 $\pi/a_0$. These points form a 'Λ' line shape, which mimics the trajectory of red dashed curve in Figure 3c. Note that the energy of the 'Λ' peak is around $E = 12$ meV, which is consistent with the energy of $|\Delta_1| = 12.4$ meV, as shown in the averaged tunneling spectrum (Figure 1d). Furthermore, the $|q|$ of the 'Λ' peak is close to that of red dashed curve, corresponding to the knowledge that the inner pocket $\boldsymbol{\delta_1}$ is close to the outer pocket $\boldsymbol{\delta_2}$ in $k$-space[11]. Therefore, the maximum-intensity points on 'Λ' line shape most likely originate from the scattering within inner CECs ($\boldsymbol{\delta_1} \leftrightarrow \boldsymbol{\delta_1}$).

In addition to the gap structure, another key characteristic of SC order parameters is the gap symmetry. Here, we proposed a PR-QPI approach, which adopts the data processing in the defect-induced bound states QPI approach (Supporting Information VI.A)[33,41-44]. The PR-QPI signals at positive and negative energies are defined as

$$g_{pr}(\mathbf{q}, +E) = |g(\mathbf{q}, +E)|\cos\left(\theta_{\mathbf{q},+E} - \theta_{\mathbf{q},-E}\right) \quad (1)$$

$$g_{pr}(\mathbf{q}, -E) = |g(\mathbf{q}, -E)| \quad (2)$$

Provided that the scattering potential is non-magnetic and relatively weak, $g_{pr}(\mathbf{q}, +E)$ should be negative for the gap-sign-changing scattering and yet positive for the gap-sign-preserving scattering (see theoretical basis of our PR-QPI approach in Supporting Information VI.B). Accordingly, if the gap signs on $\delta_2$ and $\delta_1$ are opposite, we are supposed to see negative-signal arcs in $g_{pr}(\mathbf{q},+E)$ at the energy within $|\Delta_2^{min}|$ and $|\Delta_2^{max}|$, whereas $g_{pr}(\mathbf{q},-E)$ is always positive according to the definition.

Figure 4 shows the typical FT-QPI patterns $|g(\boldsymbol{q},+E)|$ and corresponding PR-QPI patterns $g_{pr}(\boldsymbol{q},+E)$ (see Figure S12 in Supporting Information VI.C for full PR-QPI patterns). We firstly focus on the pattern at $E$ = 18 meV, i.e. the gap maximum. One can see that the most contour of $\boldsymbol{q_1}$ ring (the central scattering ring in Figure 4a) shows negative signals in the corresponding PR-$\boldsymbol{q_1}$ pattern (blue pixels of the central scattering ring in Figure 4e), indicating the sign reversal gap symmetry in 1-UC Fe(Se,Te). In addition, there are also positive signals (red pixels, albeit more fragile) inside and outside the blue pixels in the PR-$\boldsymbol{q_1}$ pattern. Figure 4i and 4j show schematics of 'bonding-antibonding' $s_\pm$-wave pairing state and 'quasi-nodeless' $d$-wave pairing state in folded BZ, respectively. There are three components of $\boldsymbol{q_1}$ pattern along the $\Gamma\widetilde{M}$ direction, consisting of $\boldsymbol{\delta_1} \leftrightarrow \boldsymbol{\delta_1}$ scattering, $\boldsymbol{\delta_1} \leftrightarrow \boldsymbol{\delta_2}$ scattering, and $\boldsymbol{\delta_2} \leftrightarrow \boldsymbol{\delta_2}$ scattering, as shown by the three arrows in Figure 4i and 4j. For both 'bonding-antibonding' $s_\pm$-wave and 'quasi-nodeless' $d$-wave, $\boldsymbol{\delta_1} \leftrightarrow \boldsymbol{\delta_1}$ and $\boldsymbol{\delta_2} \leftrightarrow \boldsymbol{\delta_2}$ scatterings are gap-sign-preserving (red arrows) and $\boldsymbol{\delta_1} \leftrightarrow \boldsymbol{\delta_2}$ scattering is gap-sign-changing (blue arrow), along the $\Gamma\widetilde{M}$ direction. Therefore, the red-blue-red pixels of PR-$\boldsymbol{q_1}$ pattern along $\Gamma\widetilde{M}$ directions (Figure 4e) could not distinguish the $s_\pm$-wave from $d$-wave. Nevertheless, the $\boldsymbol{\delta_2} \leftrightarrow \boldsymbol{\delta_2}$ scattering dominates the $\boldsymbol{q_1}$ pattern along the $\Gamma\widetilde{X}$ direction, which is gap-sign-preserving for $s_\pm$-wave (labeled as $\boldsymbol{q_1^{\Gamma\widetilde{X}}}$ in Figure 4i) and yet gap-sign-changing for $d$-wave (Figure 4j). The simulated PR-$\boldsymbol{q_1}$ patterns also show positive signals along $\Gamma\widetilde{X}$ direction for $s_\pm$-wave (Figure 4k), and yet negative signals along $\Gamma\widetilde{X}$ direction for $d$-wave (Figure 4l). In Figure 4e, the red pixels dominate the PR-$\boldsymbol{q_1}$ pattern along $\Gamma\widetilde{X}$ direction, accordingly preferring 'bonding-antibonding' $s_\pm$-wave to 'quasi-nodeless' $d$-wave. We also analyze PR-$\boldsymbol{q_2}$ scattering pattern at $E$ = 18 meV in Supporting Information VI.C.

Then, we concentrate on the evolution of PR-QPI patterns as energy varies, which is supposed to follow the same trend as FT-QPI patterns (Figure 4a-d). As expected, the region where negative signals occupy gradually shrinks as energy decreases from 18 meV to 12 meV (Figure 4e-g), due to the contraction of the banana-shape CECs of $\boldsymbol{\delta_2}$. Especially, at $E$ = 12 meV, the negative signals only aggregate in small regions centered in $\Gamma\widetilde{M}$ directions (Figure 4g). Eventually, the positive signals become conspicuous, completely covering the $\boldsymbol{q_1}$ ring at $E$ = 10 meV (Figure 4h), which is in accordance with the obtained gap minimum $|\Delta_2^{min}| \approx 10$ meV. Furthermore, the evolution of our PR-QPI signals from negative to positive as energy decreases can undoubtedly corroborate the gap sign change and the non-magnetic scattering potential simultaneously. For magnetic scattering potential, the gap-sign-changing scattering would present positive signals in PR-QPI patterns, while the gap-sign-preserving scattering would present negative signals (Supporting Information VI.B). If the gap sign does not change in the two pockets, the PR-QPI signals would remain positive for non-magnetic scattering potential and negative for magnetic scattering potential as energy varies, both contradicting our observation. Thus, the gap sign must change between the inner pocket $\boldsymbol{\delta_1}$ and the outer pocket $\boldsymbol{\delta_2}$. In this case, while the gap-sign-changing scattering ($\boldsymbol{\delta_1} \leftrightarrow \boldsymbol{\delta_2}$) dominates the PR-QPI patterns at higher energy, the gap-sign-preserving scattering ($\boldsymbol{\delta_1} \leftrightarrow \boldsymbol{\delta_1}$) would control the PR-QPI patterns when energy decreases to $|\Delta_2^{min}|$, due to the contraction and eventual disappearance of $\boldsymbol{\delta_2}$. Accordingly, for magnetic scattering potential, the PR-QPI signals are supposed to change from positive to negative as energy decreases, which is totally opposite of our observation. Therefore, only the case of non-magnetic scattering potential and gap sign change is consistent with the evolution of our PR-QPI signals from negative to positive as energy decreasing.

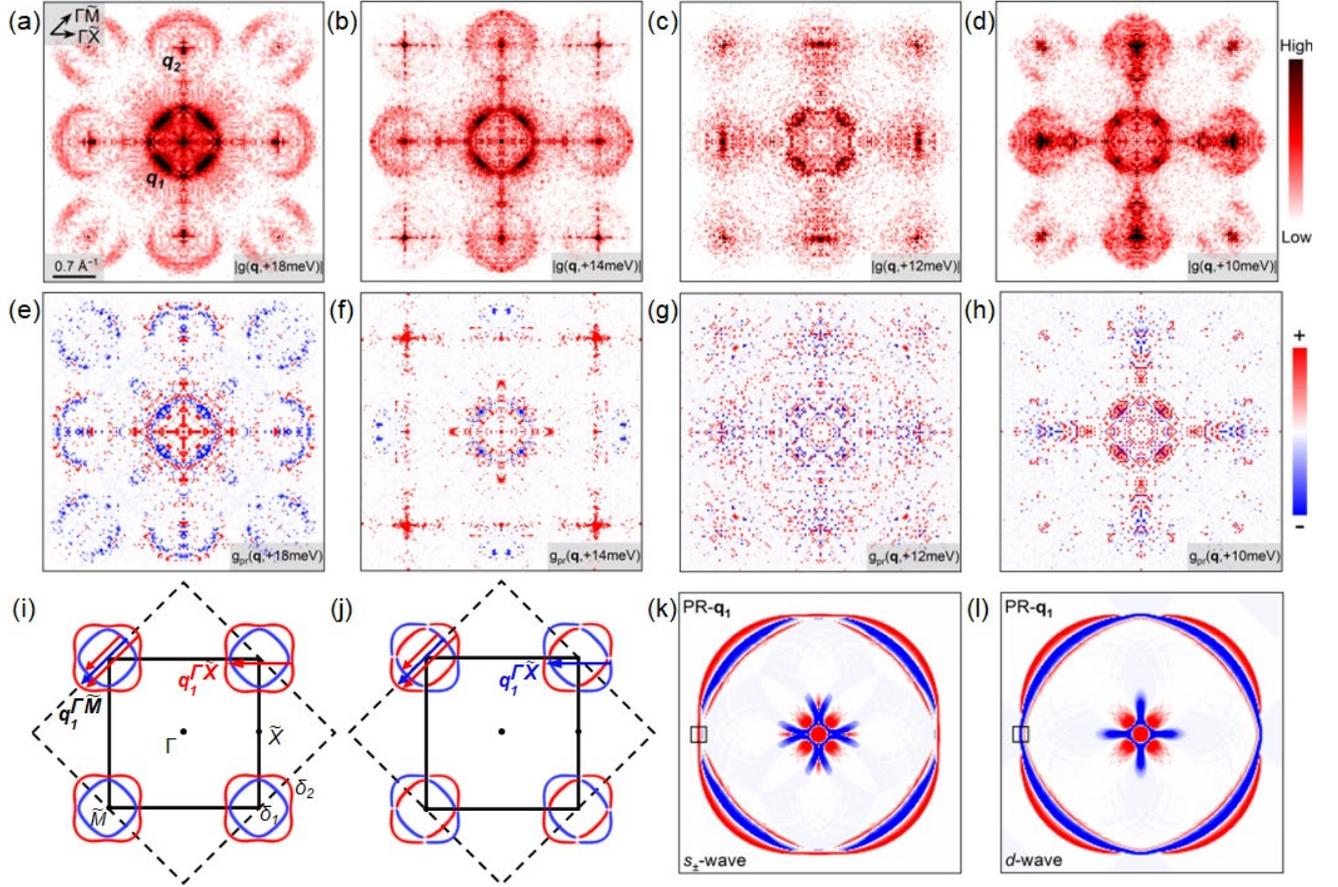

**Figure 4. Sign-reversal SC gaps revealed by PR-QPI.** The typical FT-QPI patterns $|g(\boldsymbol{q},+E)|$ (a-d) and corresponding PR-QPI patterns $g_{pr}(\boldsymbol{q},+E)$ (e-h). As the size of CECs of $\boldsymbol{\delta_2}$ gradually shrinks with decreasing energy from 18 meV to 10 meV, the gap-sign-changing scattering signals (blue pixels of the central scattering ring $\boldsymbol{q_1}$ in (e-h)) constantly diminish and finally disappear. (i,j) Schematics of 'bonding-antibonding' $s_\pm$-wave pairing state and 'quasi-nodeless' $d$-wave pairing state in folded BZ, respectively. The red and blue colors of the pockets ($\boldsymbol{\delta_1}$ and $\boldsymbol{\delta_2}$) represent opposite signs of SC gaps. The red and blue arrows denote the gap-sign-preserving and gap-sign-changing scattering vectors, respectively. Note that there is no difference between 'quasi-nodeless' $d$-wave and 'bonding-antibonding' $s_\pm$-wave for scattering pattern along the $\Gamma\widetilde{M}$ direction. However, the scattering pattern along the $\Gamma\widetilde{X}$ direction labeled as $\boldsymbol{q_1^{\Gamma\widetilde{X}}}$ mainly consists of gap-sign-preserving scattering for 'bonding-antibonding' $s_\pm$-wave. Whereas, the scattering pattern along the $\Gamma\widetilde{X}$ direction is dominated by gap-sign-changing scattering for 'quasi-nodeless' $d$-wave. (k,l) Simulated phase-referenced $\boldsymbol{q_1}$ patterns according to the pairing symmetries in (i) and (j), respectively. The black box marks the signal in the $\Gamma\widetilde{X}$ direction, which shows red pixels (gap-sign-preserving scattering) for 'bonding-antibonding' $s_\pm$-wave and blue pixels (gap-sign-changing scattering) for 'quasi-nodeless' $d$-wave.

By using PR-QPI approach, we directly visualized the opposite SC gap signs on the two electron pockets, which prefer the 'bonding-antibonding' $s_\pm$-wave pairing symmetry to the 'quasi-nodeless' $d$-wave pairing symmetry, according to the PR-$\boldsymbol{q_1}$ pattern along $\Gamma\widetilde{X}$ direction. Furthermore, the anisotropic gap structure also supports $s_\pm$-wave and opposes $d$-wave. For $d$-wave pairing symmetry, since the intra-orbital inter-pocket $d_{xy} \leftrightarrow d_{xy}$ pair scattering can take advantage of the major spin fluctuations (SF) at (π, π) in the unfolded BZ (see Figure S8 in Supporting Information IV for the orbital character of FSs), the SC gap is supposed to be maximum at the tip of the elliptic-shaped pockets where the $d_{xy}$ orbital has largest weight[45]. However, this naturally generates the gap maximum in the $\Gamma\widetilde{M}$ direction (in the folded BZ) and gap minimum, albeit not zero, in the $\Gamma\widetilde{X}$ direction, where the two ellipses crossed[16], which fully conflicts with our determination of gap maximum in the $\Gamma\widetilde{X}$ direction and gap minimum in the $\Gamma\widetilde{M}$ direction (Figure 2f and Figure 3d).

In the microscopic theory of superconductivity, the formation of superconductivity not only opens a low-energy SC gap $\Delta(\boldsymbol{k})$, but also renormalizes the higher-energy band dispersion associated with the bosonic mode[39]. Recently, it is shown that the band dispersion extracted from the FT-QPI patterns in 1-UC FeSe/STO is heavily renormalized in the $\Gamma\widetilde{X}$ direction at $E$ = 37 meV, coinciding in the bosonic mode energy detected in the tunneling spectrum, whereas this dispersion renormalization is moderate in the $\Gamma\widetilde{M}$ direction[46]. This result once again confirms that the electron pairs in the $\Gamma\widetilde{X}$ direction generate a larger SC gap and a stronger band renormalization at the same time. We noticed that upon increasing the band hybridization of the two elliptic-shaped electron pockets, the interplay between intra-pocket and inter-pocket pairings will transform the system from a $d$-wave state to an $s_\pm$-wave state, where the location of the gap maximum simultaneously transforms from $\Gamma\widetilde{M}$ direction to $\Gamma\widetilde{X}$ direction[14]. This proposal supports the 'bonding-antibonding' $s_\pm$-wave scenarios as the pairing symmetry and qualitatively matches the gap structure we extracted.

Widely discussed ingredients for topological superconductors are (i) unconventional pairing in multiband systems and (ii) SOC. This work directly unravels the nature of the SC order parameter in showing that it is unconventional (with sign change). Overall, both the gap sign change and the gap structure in 1-UC Fe(Se,Te)/STO are well consistent with the 'bonding-antibonding' $s_\pm$-wave paring symmetry, indicating a unified picture of pairing in FeBSs. The 'bonding-antibonding' $s_\pm$- state can be driven by SF in conjunction with hybridization of the electron pockets from SOC, thus our work also gives valuable input for further understanding of whether and how topological superconductivity can be realized in FeBSs.

## ■ ASSOCIATED CONTENT

**Supporting Information**
Materials synthesis and experimental method, more information about the 1-UC Fe(Se,Te)/STO and the Dynes model fitting, FT-QPI data, the relationship between orbital characters and scattering patterns, two approaches to extract gap structure, the PR-QPI approach

## ■ AUTHOR INFORMATION

**Author contributions**
J.W. conceived and instructed the research. Y.L. and C.C. prepared the samples and carried out the STM/STS experiments. Y.L., D.S., X.X., and T.W. analyzed the experimental data. D.S. and A.K. performed the theoretical analysis and calculations. Y.L., D.S., and J.W. wrote the manuscript with comments from all authors.

**Notes**
The authors declare no competing financial interest.

■ ACKNOWLEDGMENTS

The authors acknowledge the discussions with Brian M. Andersen and Peter J. Hirschfeld. This work was financially supported by Beijing Natural Science Foundation (No. Z180010), National Natural Science Foundation of China (No.11888101), National Key R&D Program of China (No. 2018YFA0305604), and Strategic Priority Research Program of Chinese Academy of Sciences (No. XDB28000000).

Supporting Information for

# Anisotropic gap structure and sign reversal symmetry in monolayer Fe(Se,Te)

Yu Li,[†] Dingyu Shen,[†] Andreas Kreisel,[‡] Cheng Chen,[†] Tianheng Wei,[†] Xiaotong Xu[†] and Jian Wang[†,§,∥,#,*]

[†]International Center for Quantum Materials, School of Physics, Peking University, Beijing 100871, China.
[‡]Institut für Theoretische Physik, Universität Leipzig, D-04103 Leipzig, Germany.
[§]Collaborative Innovation Center of Quantum Matter, Beijing 100871, China.
[∥]CAS Center for Excellence in Topological Quantum Computation, University of Chinese Academy of Sciences, Beijing 100190, China.
[#]Beijing Academy of Quantum Information Sciences, Beijing 100193, China.

*Corresponding author.
E-mail: jianwangphysics@pku.edu.cn.

**CONTENTS**

I. Materials Synthesis and Experimental Method
II. More Information about the One Unit-cell (1-UC) Fe(Se,Te)/STO and Dynes Model Fitting
III. Fourier Transform Quasiparticle Interference Data
IV. Relationship between Orbital Characters and Scattering Patterns
V. Two Approaches to Extract Gap Structure
VI. Phase-referenced Quasiparticle Interference Approach
Figures S1-S12
References [1]-[35]

# I. Materials Synthesis and Experimental Method

## A. Materials Synthesis

Our experiments were performed in an ultrahigh-vacuum (~$2\times10^{-10}$ mbar) MBE-STM combined system (Scienta Omicron). The Nb-doped $SrTiO_3$(001) (wt 0.7 %) substrates were thermally boiled in 90 °C deionized water for 50 minutes and then chemically etched in 10% HCl for 45 minutes. Followed by Se-flux method in a MBE chamber, the substrates finally obtain the atomically flat $TiO_2$-terminated surface. The 1-UC Fe(Se,Te) films were grown by co-evaporating high-purity Fe (99.994%) Se (99.999%) and Te (99.999%), with the substrates held at 340 °C. Then the as-grown 1-UC Fe(Se,Te) films were annealed at 380 °C for 3 hours.

## B. Scanning Tunneling Microscopy/ Spectroscopy (STM/S) Measurements

All STM/STS data were acquired in the *in-situ* STM chamber with a polycrystalline Pt/Ir tip by using the standard lock-in technique. The modulation voltage was $V_{mod}$ = 0.5 mV at 1.7699 kHz. The setup of STM topographic images is $V$ = 0.1 V, $I$ = 0.5 nA unless specified. And the setup of all STS measurements is $V$ = 0.04 V, $I$ = 2.5 nA for tunneling spectra and differential conductance mapping, i.e. quasiparticle interference (QPI). All STM/STS measurements were taken at 4.3 K. The energy resolution of all STS measurements is $\Delta E = \sqrt{(3.5k_B T)^2 + (2.5eV_{mod})^2} = 1.8$ meV, where $k_B$ is the Boltzmann constant.

Fig. S1a and S1b show the data process of a typical tunneling spectrum and an averaged (over 180 spectra) tunneling spectrum, respectively. The raw d$I$/d$V$ spectrum (top panel of Fig. S1a and S1b) was normalized (middle panel of Fig. S1a and S1b) by dividing it by its polynomial background, which was acquired by a cubic fitting to the raw d$I$/d$V$ spectrum for $|V| \geq 30$ mV. For the typical tunneling spectrum, one could directly determine the two SC gaps of $\Delta_1$ = -11.0 meV / 11.0 meV and $\Delta_2$ = -18.2 meV / 18.8 meV, by locating the zero points in the second derivative of conductance (d$^2I$/d$V^2$), as shown in the bottom panel of Fig. S1a. For the averaged tunneling spectrum, while the SC gap of $\Delta_2$ = -18 / 18 meV can be determined by the zero points in d$^2I$/d$V^2$ spectrum, the SC gap of $\Delta_1$ = 12.4 meV could be estimated by the local minimum in d$^2I$/d$V^2$ spectrum (bottom panel of Fig. S1b). Originating from the band edge effect[1], the tunneling spectrum is quite asymmetric with density of states much higher at positive bias voltage than the negative bias side, which holds even after the normalization (middle panel of Fig. S1a and S1b). Accordingly, the particle-hole symmetrization (top panel of Fig. S1c and S1d) was performed by averaging the negative bias side and the positive bias side of the normalized d$I$/d$V$ spectrum. Then, according to the zero points in d$^2I$/d$V^2$ spectrum (bottom panel of Fig. S1c and S1d), one could determine the SC gaps of $|\Delta_1|$ = 11.0 meV and $|\Delta_2|$ = 18.4 meV in the typical tunneling spectrum and SC gap of $|\Delta_2|$ = 18.0 meV in the averaged tunneling spectrum. By locating the local maximum or minimum in d$^2I$/d$V^2$, one could estimate the SC gap of $|\Delta_1|$ = 12.4 meV in the averaged tunneling spectrum. In addition, the tunneling spectrum displays an obvious hump feature outside the SC gaps (orange area in Fig. S1a), which implies the bosonic mode and is consistent with our previous results[2]. Consequently, we picked the bosonic-coupling-unmodified bias region [−25,25] mV for the Dynes model fitting (Figure 1c and 1d).

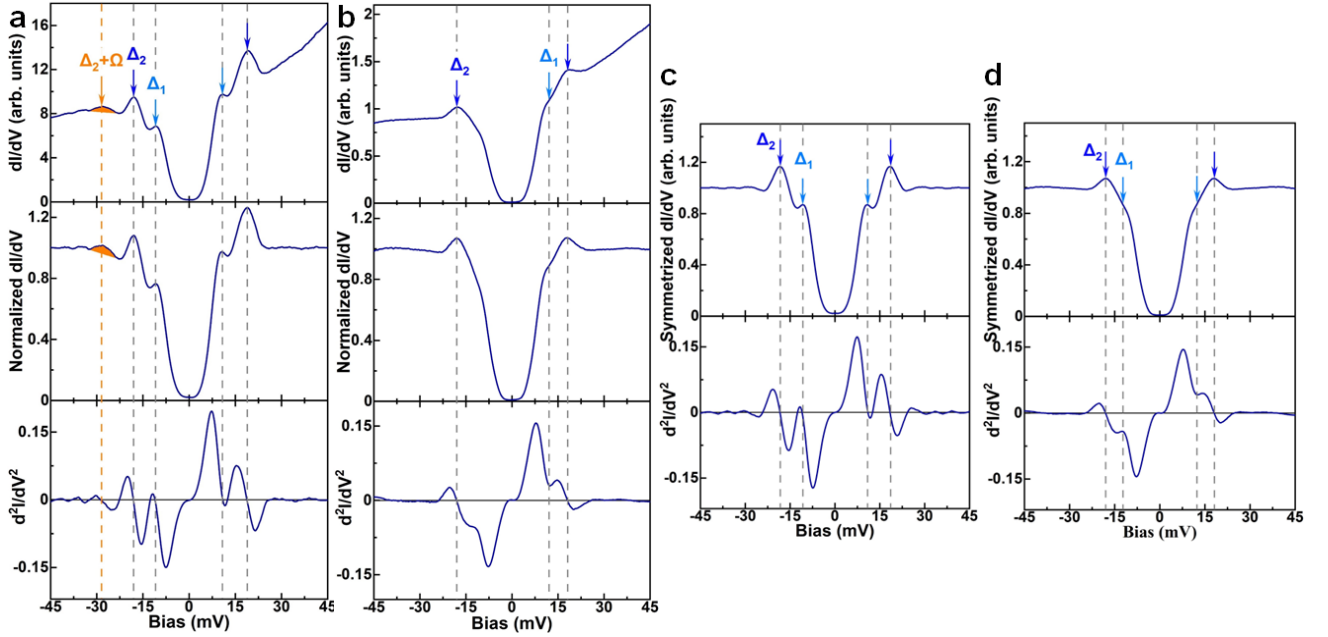

**Figure S1. Data process of STS spectra. a**, Typical tunneling spectra showing raw $dI/dV$ (top panel), normalized $dI/dV$ (middle panel), and second derivative of conductance $d^2I/dV^2$ (bottom panel) at 4.3 K. The gray and orange dashed lines show the energies of $\Delta_1$ = -11.0 meV / 11.0 meV, $\Delta_2$ = -18.2 meV / 18.8 meV and $\Delta_2 + \Omega$ = -28.2 meV, and the orange areas highlight the hump features developing from the bosonic mode $\Omega$. **b**, An averaged (over 180 spectra) tunneling spectrum at 4.3 K, followed the same data processing procedure as **a**. The gray dashed lines show the energies of $\Delta_1$ = 12.4 meV and $\Delta_2$ = -18 meV / 18 meV. **c**, The typical $dI/dV$ spectrum after normalization and symmetrization (top panel) and corresponding $d^2I/dV^2$ spectrum (bottom panel). The gray dashed lines show the energies of $|\Delta_1|$ = 11.0 meV and $|\Delta_2|$ = 18.4 meV. **d**, The averaged $dI/dV$ spectrum after normalization and symmetrization (top panel) and corresponding $d^2I/dV^2$ spectrum (bottom panel). The gray dashed lines show the energies of $|\Delta_1|$ = 12.4 meV and $|\Delta_2|$ = 18.0 meV.

## II. More Information about the One Unit-cell (1-UC) Fe(Se,Te)/STO and Dynes Model Fitting
### A. The Te/Se Ratio and Native Defects in 1-UC Fe(Se,Te)/STO

We used the thickness of the $2^{nd}$-UC Fe(Se,Te) film to estimate the Se content (Fig. S2a and S2b). Empirically, the thickness of the $2^{nd}$-UC Fe(Se,Te) is about 0.59 nm when Se concentration equals 50%[3]. The main defect in 1-UC Fe(Se,Te)/STO is the dumbbell-like impurity, consisting of two adjacent top-layer Se/Te atoms (Fig. S2c). Since the height of the bright site is only 34 ± 5 pm (Fig. S2d), we interpret it as the Fe-centered vacancy in the underlying layer rather than extra Se/Te atoms, which is widely accepted in 1-UC FeSe[4]. The spectra measured around the defect show that the coherence peak of the SC gap at negative bias side is remarkably enhanced, while the zero-bias conductance remains at zero (Fig. S2e). Commonly, when the defect-induced scattering potential is weak, the bound state lies close to the gap edge[5]. Furthermore, the same type of defects in $(Li_{1-x}Fe_x)OHFe_{1-y}Zn_ySe$ is proved to be nonmagnetic because of the non-shift of the peak energy under the magnetic field of 11 T[6]. Therefore, the dominant defects in 1-UC Fe(Se,Te) likely have weak and non-magnetic scattering potential, which is favorable to explicitly judging the gap-sign issue in 1-UC Fe(Se,Te) by the PR-QPI approach.

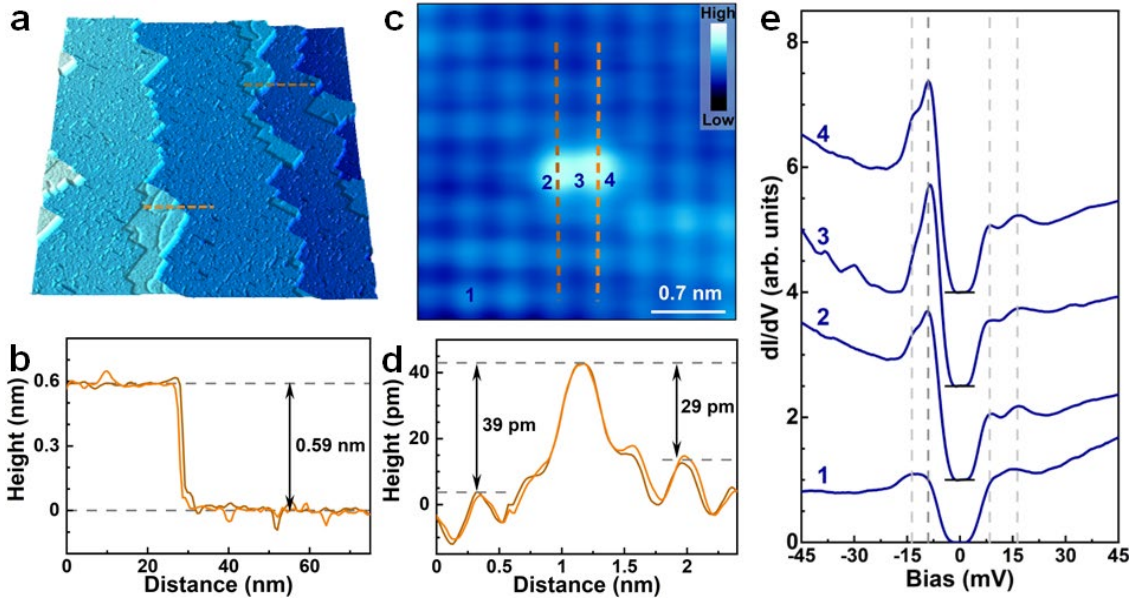

**Figure S2. 1-UC Fe(Se,Te)/STO Characterization. a**, The large-scale STM topography of 1-2 UC Fe(Se,Te)/STO (300×300 nm$^2$, $V$ = 1 V, $I$ = 0.5 nA), where the lighter color means the higher height. Apart from the terraces inherited from the STO substrate, the 2$^{nd}$-UC Fe(Se,Te) film normally grows on the edge of the lower terrace. **b**, Profile taken along the orange and brown curves in **a**, respectively, both showing the thickness of the 2$^{nd}$-UC Fe(Se,Te) film is 0.59 nm. Since the thickness of the 2$^{nd}$-UC Fe(Se,Te) film changes with the ratio of Se/Te, height of 0.59 nm indicates the Se concentration is around 50%[3]. **c**, Topographic images of a dumbbell-like defect arising from Fe vacancy (3×3 nm$^2$, $V$ = 0.1 V, $I$ = 0.5 nA). **d**, Profile taken along the orange and brown curves in **a**, respectively, showing the height of the bright site is 34 ± 5 pm. **e**, Tunneling spectra ($V$ = 0.04 V, $I$ = 2.5 nA, vertically offset for clarity) obtained at the positions as numbered in **c**. The gray dashed lines indicate the two pairs of SC coherence peaks. The bound states are presented as the enhanced DOSs at the negative bias side near the gap edge.

## B. Comparison of Different Types of Dynes Model

Figure S3 and Figure S4 show the Dynes model fitting results of the individual tunneling spectrum and the averaged tunneling spectrum, respectively. We compared four different types of Dynes model and defined a quantity $e = \sqrt{\sum \delta y^2}$ to describe the goodness of every fit, where $\delta y$ is the residual error. The simplest form is the original Dynes model[7,8] also named one-band isotropic Dynes model.

$$\frac{dI}{dV}(eV) = N(E_F) \frac{1}{k_B T} \int_{-\infty}^{+\infty} dE \, Re\left[\frac{|E - i\Gamma|}{\sqrt{(E - i\Gamma)^2 - \Delta^2}}\right] \cosh^{-2} \frac{E + eV}{2k_B T} \quad (1)$$

Here, $N(E_F)$ is the local density of states (DOSs) at $E_F$, $\Gamma$ is the scattering rate, and $k_B$ is the Boltzmann constant. Obviously, the one-band isotropic Dynes model cannot fit the tunneling spectrum (Fig. S3a) with an abnormally large $e$ = 1.13 and only gives a single superconducting (SC) gap $|\Delta|$ = 8.58 meV.

The fitting incorporating anisotropic gap structure (Fig. S3b) has a significant improvement compared to the isotropic gap fitting, where $e$ dramatically decreases to 0.42 and the fitting parameter $|\Delta_2^{max}|$ = 18.17 meV is comparable to the energy of the large coherence peaks (18.5 meV). The fitting functions are given by

$$\frac{dI}{dV} = N(E_F) \frac{1}{2\Pi} \frac{1}{k_B T} \int_{-\infty}^{+\infty} dE \int_0^{2\Pi} d\theta \, Re\left[\frac{|E - i\Gamma|}{\sqrt{(E - i\Gamma)^2 - \Delta(\theta)^2}}\right] \cosh^{-2} \frac{E + eV}{2k_B T} \quad (2)$$

$$\Delta(\theta) = \Delta^{max}[1 - p_1(1 - \cos 4\theta) - p_2(1 - \cos 2\theta)] \quad (3)$$

where the anisotropic function was adopted on account of its consistency with angle-resolved photoemission spectroscopy (ARPES) measurements of the angular-dependent $\Delta(\theta)$ in 1-UC FeSe[9], and the ability to give the second maximum at $\theta$ = 90° / 270°. This second maximum indeed corresponds to the small SC gap $|\Delta_1^{max}|$ = 9.09 meV. However, it is moderately less than the energy of the small coherence peak (11 meV), also shown as the observable deviation between the fitting curve and the tunneling spectrum in Fig. S3b.

From a theoretical perspective, the band folding and hybridization in BZ of 2-Fe unit cell shall naturally recombine the two elliptic-shaped electron pockets into the inner and outer pockets (Figure 2b). Then the newly formed inner and outer pockets will possess distinct gap functions. Accordingly, we used the summation of two Dynes functions with different weights ($w$ vs. $1-w$) to represent the two-band model, given by

$$\frac{dI}{dV} = N(E_F)\left[w\frac{dI_1}{dV} + (1-w)\frac{dI_2}{dV}\right] \quad (4)$$

Though the two-band isotropic Dynes model (Fig. S3c) fits the tunneling spectrum well ($e$ = 0.27), the obtained parameters of $|\Delta_1|$ = 7.87 meV and $|\Delta_2|$ = 16.46 meV are unreasonably small, coming from the unusually large scattering rates of $\Gamma_1$ = 2.41 meV and $\Gamma_2$ = 1.31 meV.

After including gap anisotropy, two-band anisotropic Dynes model (Fig. S3d) displays the best fitting performance with lowest $e$ = 0.26 and gives the most reasonable fitting parameter of $|\Delta_1^{max}|$ = 9.55 meV, $|\Delta_2^{max}|$ = 18.13 meV, $\Gamma_1$ = 0.51 meV, and $\Gamma_2$ = 0.79 meV, where the anisotropic function is given by

$$\Delta_{1/2}(\theta) = \Delta_{1/2}^{max}[1 - p_{1/2}(1 - \cos 4\theta)] \quad (5)$$

The obtained gap values show moderate decrease compared to those determined by $d^2I/dV^2$, since the Dynes model considers the broadening stemming from finite-lifetime effects of the quasiparticles at the gap edge[7,8].

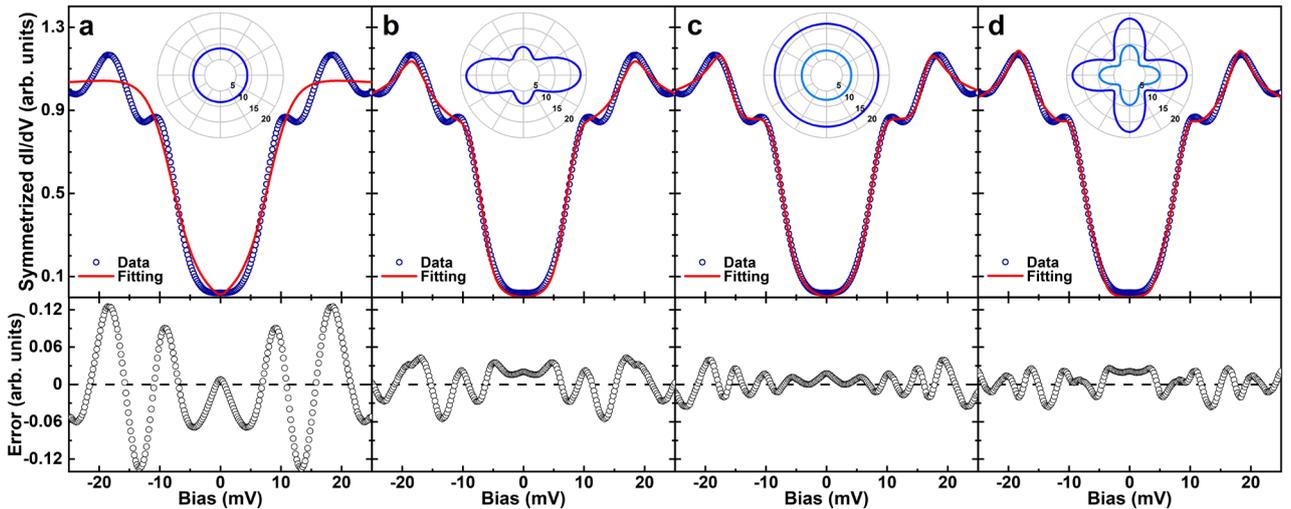

**Figure S3. Fittings of the individual tunneling spectrum with different types of Dynes model.** In the top panel, the blue hollow dots represent the symmetrized d$I$/d$V$ spectrum and the red solid line is the theoretical fittings by different types of Dynes model. Inset shows the gap functions used in each Dynes model. Bottom panel shows the residual error defined as the symmetrized d$I$/d$V$ spectrum subtracting the fitted curve. **a**, One-band isotropic Dynes model fitting. Fitting parameters: $|\Delta|$ = 8.58 meV, $\Gamma$ = 5.12 meV. Fitting error: $e$ = 1.13. **b**, One-band anisotropic

Dynes model fitting. $|\Delta_1|$ = 9.09 meV, $|\Delta_2|$ = 18.17 meV, $\Gamma$ = 1.28 meV, $e$ = 0.42. **c**, Two-band isotropic Dynes model fitting. $|\Delta_1|$ = 7.87 meV, $|\Delta_2|$ = 16.46 meV, $\Gamma_1$ = 2.41 meV, $\Gamma_2$ = 1.31 meV, $e$ = 0.27. **d**, Two-band anisotropic Dynes model fitting. $|\Delta_1|$ = 9.55 meV, $|\Delta_2|$ = 18.13 meV, $\Gamma_1$ = 0.51 meV, $\Gamma_2$ = 0.79 meV, $e$ = 0.26.

The fittings of the averaged tunneling spectrum further confirms that the two-band isotropic Dynes model provides unreasonable parameters, and two-band anisotropic Dynes model gives the best fitting (Fig. S4). One tiny difference in this case is that the one-band anisotropic Dynes model shows almost the same results with two-band anisotropic Dynes model. Overall, whether one-band or two-band, incorporating anisotropic gap functions dramatically promotes the performance of Dynes model, revealing the significance of SC gap anisotropy.

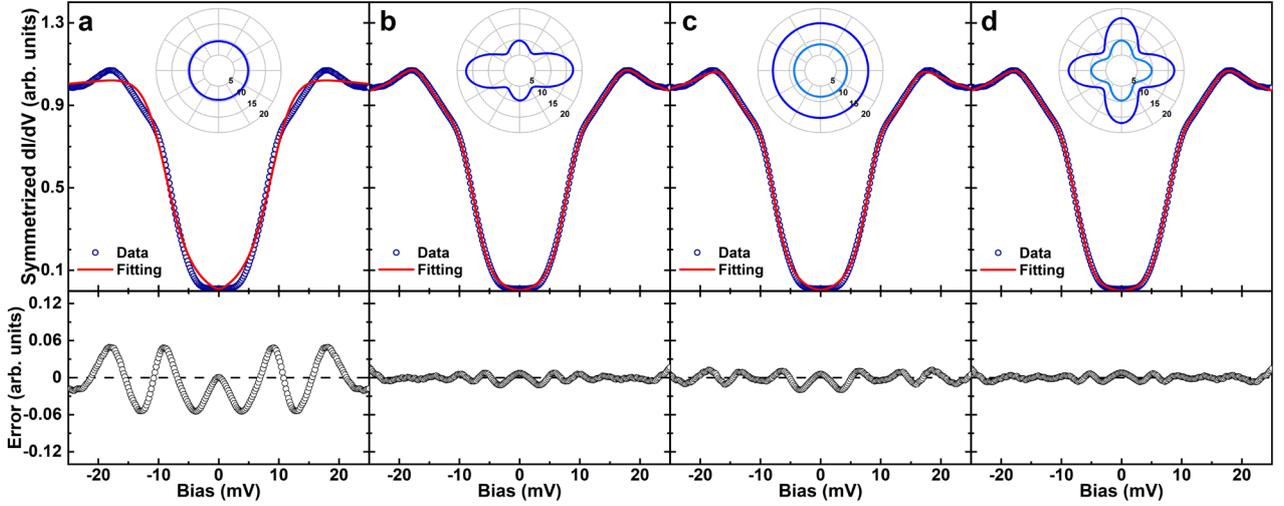

**Figure S4. Fittings of the averaged tunneling spectrum with different types of Dynes model.** Same data processing as Fig. S3 but for the averaged spectrum. **a**, One-band isotropic Dynes model fitting. Fitting parameters: $|\Delta|$ = 9.44 meV, $\Gamma$ = 4.77 meV. Fitting error: $e$ = 0.52. **b**, One-band anisotropic Dynes model fitting. $|\Delta_1|$ = 9.64 meV, $|\Delta_2|$ = 17.12 meV, $\Gamma$ = 2.23 meV, $e$ = 0.08. **c**, Two-band isotropic Dynes model fitting. $|\Delta_1|$ = 8.44 meV, $|\Delta_2|$ = 15.26 meV, $\Gamma_1$ = 2.86 meV, $\Gamma_2$ = 2.79 meV, $e$ = 0.13. **d**, Two-band anisotropic Dynes model fitting. $|\Delta_1|$ = 9.66 meV, $|\Delta_2|$ = 16.91 meV, $\Gamma_1$ = 1.54 meV, $\Gamma_2$ = 2.52 meV, $e$ = 0.07.

### III. Fourier Transform (FT-) Quasiparticle Interference (QPI) Data

Differential conductance mappings $g(r, E)$ were measured individually for each pair of positive and negative energies ($\pm E$), and each scanning took about 1000 minutes. Before the Fourier transform (FT-), the $g(r, E)$ was corrected by the Lawler-Fujita algorithm[10] to reduce the distortions of the non-orthogonality in the $x/y$ axes (Supporting Information III.A). The FT-QPI pattern was further symmetrized in order to increase the signal-to-noise ratio. Then we suppressed the intensity of very small scattering vectors around the center, since they stem from long range spatial variations of the surface and from randomly distributed defects, which occasionally obscure the visibility of the central scattering contour (Supporting Information III.B).

### A. Drift Correction by Lawler-Fujita Algorithm

We used Lawler-Fujita Algorithm to correct the distortion of $g(r, E)$[10], which is inferred from the simultaneously acquired topographic images. Due to the hysteresis and mechanical creep of piezoelectric actuator as well as the thermal drift, STM has an inevitable discrepancy between the STM controller's model of the probe tip's location and the probe tip's true location over the surface[11]. Fig. S5a and S5c typify a topographic image and the corresponding FT pattern, respectively. The inset in Fig. S5c shows the Bragg spot of the top Se lattice with noticeable noise. Furthermore, the upper and lower Bragg points obviously are not along the same vertical line, leading to non-orthogonality of the $x/y$ axes. Figure S5b and S5d show the topographic image and the corresponding FT pattern after correction. Noise around the Bragg spot decreases dramatically and the line connecting upper and lower (left and right) Bragg points is strictly vertical (horizontal) passing through the center. The latter enables us to directly symmetrize the FT-QPI patterns (Fig. S6b).

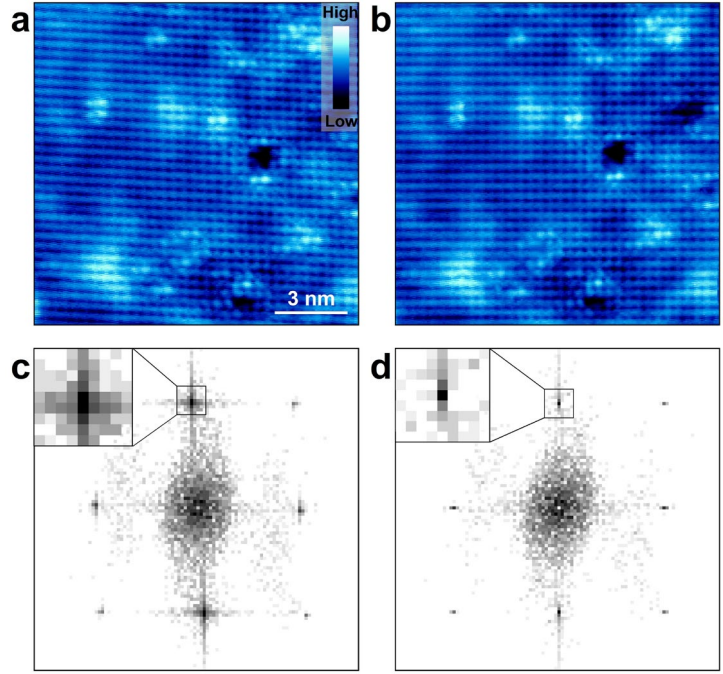

**Figure S5. Topographic images and corresponding FT patterns. a,b**, Topographic images before and after the correction (13×13 nm$^2$, $V$ = 0.04 V, $I$ = 2.5 nA), respectively. **c,d**, FT patterns of **a** and **b**, respectively. The insets are the zoom-in views of the Bragg spots.

**B. FT-QPI Data Processing Steps**

After correction of $g(r,E)$, the obtained raw $|g(q,E)|$ = |FT-$g(r,E)$| (Fig. S6a) follows successive steps to optimize the signal-to-noise ratio. Firstly, we take advantage of the mirror symmetry axes of the Fermi surfaces (FSs) geometry in 1-UC Fe(Se,Te), which are horizontal and vertical in the 1-Fe BZ and diagonal in the 2-Fe BZ. Since we have already corrected the distortion of the non-orthogonality in the $x/y$ axes (Fig. S5), we can directly symmetrize $|g(q,E)|$ via diagonal axes. This step strikingly improves the visibility of the scattering rings (Fig. S6b). Secondly, we further implemented a four-fold rotational ($C_4$) symmetrization (Fig. S6c), which is commonly used for FT-QPI data processing in 1-UC FeSe[12,13]. Previous studies reported that the pronounced nematic order present in multilayer FeSe is suppressed in 1-UC FeSe and there is no indication for a $C_4$ symmetry breaking of the electronic structure[14]. Even if the specific defects (Fig. S2c) may have a $C_2$ scattering potential ($V_{xz,xz} \neq V_{yz,yz}$), which mainly tunes the intensity of the scattering rings without modifying the scattering wave vectors, it does not break the $C_4$ symmetry of the electronic structure in 1-UC

Fe(Se,Te). Thus, we used the $C_4$ symmetrization. The last step is a Gaussian core subtraction of very small scattering vectors around the center (Fig. S6d) as $|g_{new}(\boldsymbol{q},E)| = |g_{raw}(\boldsymbol{q},E)|\,[1 - A *$ *Gaussian* $(\boldsymbol{q} = (0, 0), \sigma)]$, which restores the clarity of $\boldsymbol{q_1}$ scattering ring. The central peak around $\boldsymbol{q} = (0,0)$ chiefly develops from the randomly scattering defects and the long-range variations of the surface. Consequently, the Gaussian core subtraction corresponds to a long wavelength filter in real space, which does not affect the scattering signals of interest[15,16]. The parameters A and σ were adjusted slightly to make $\boldsymbol{q_1}$ scattering rings at different energies as clear as possible (Fig. S7).

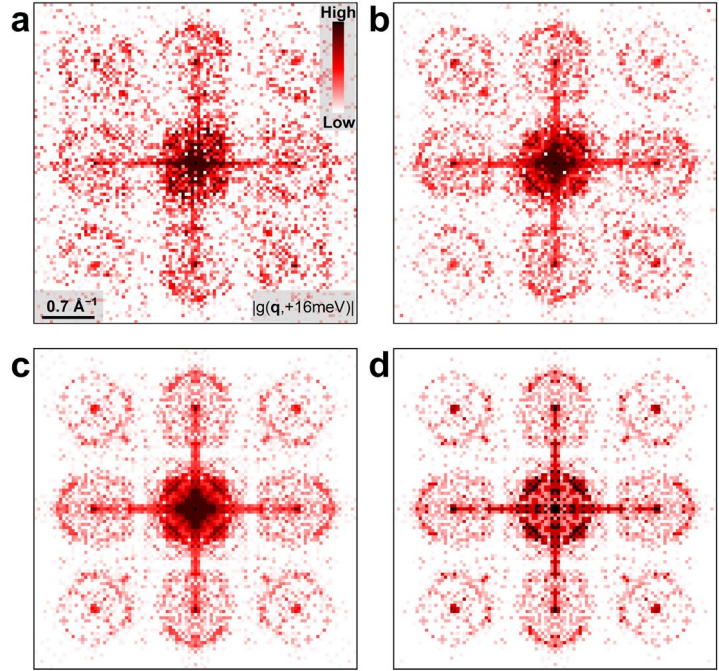

**Figure S6. Exemplifying the FT-QPI pattern processing steps. a**, Raw $|g(\boldsymbol{q}, E = 16$ meV$)|$ as obtained from distortion corrected $g(\boldsymbol{r}, E = 16$ meV$)$ shown in Figure 2a. **b**, Mirror symmetrized $|g(\boldsymbol{q}, E = 16$ meV$)|$. **c**, Mirror symmetrized and $C_4$ symmetrized $|g(\boldsymbol{q}, E = 16$ meV$)|$. **d**, Mirror symmetrized, $C_4$ symmetrized, and Gaussian core subtracted $|g(\boldsymbol{q}, E = 16$ meV$)|$.

## C. FT-QPI Data

Limited by the duration time of scanning tunneling spectroscopy before next time liquid helium filling, $g(\boldsymbol{r}, E)$ at different $E$ were inevitably taken at several different filed of views (FOVs). Thus, it is necessary to normalize the intensity $|g(\boldsymbol{q}, E)|$ at different $E$ before plotting together (Figure 2f). Besides, the variation of electronic structure between different areas in 1-UC Fe(Se,Te) film might induce the deviation of band dispersion at energy $E = 11$ meV and $E = 15$ meV (orange dashed curve in Figure 3c).

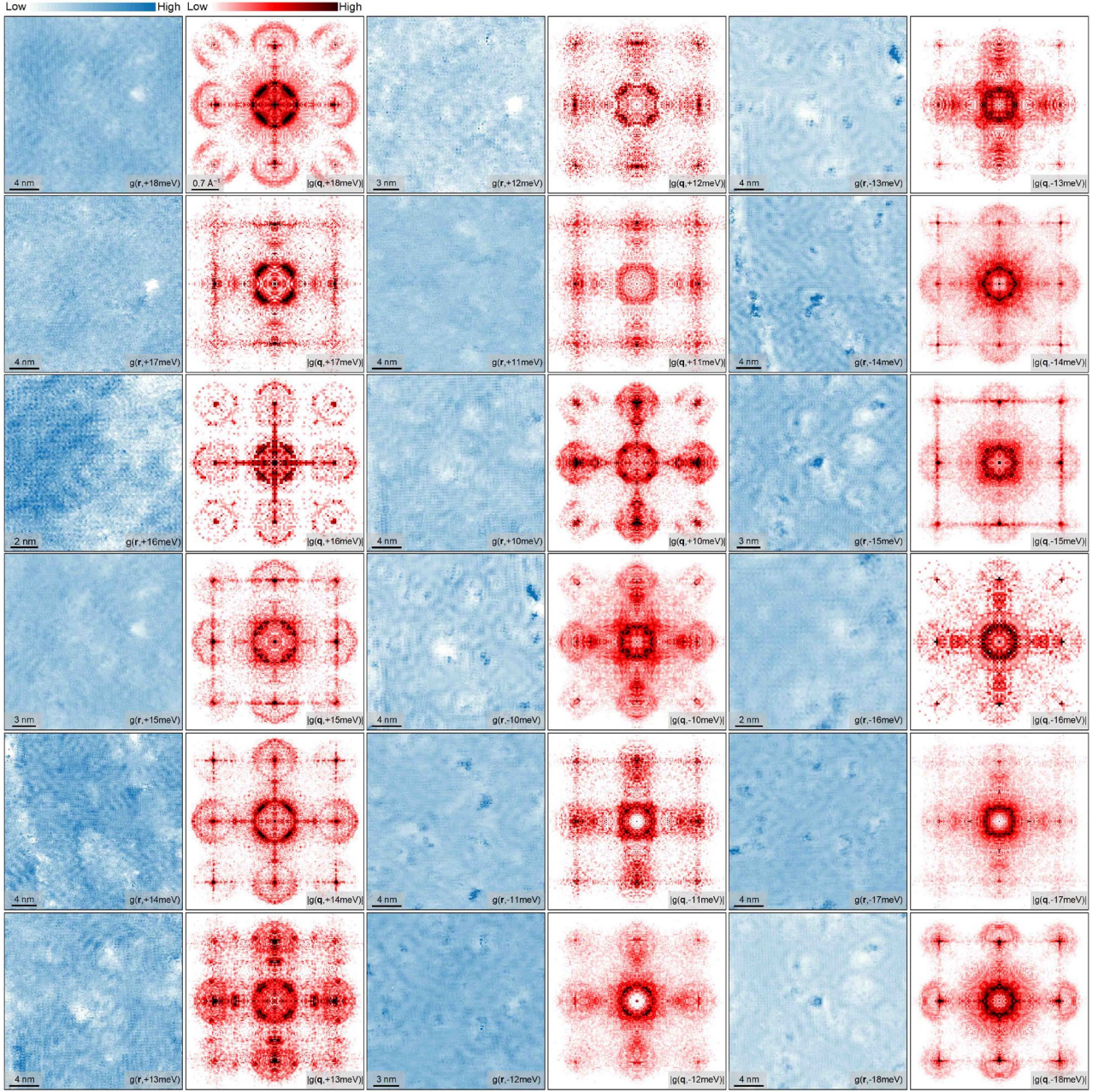

**Figure S7. Differential conductance mappings $g(r, E)$ and corresponding FT-QPI patterns $|g(q, E)|$.** Differential conductance mappings at different $E$ were taken at different FOVs. $g(r, E = \pm 16$ meV) were taken in a $13 \times 13$ nm$^2$ area. $g(r, E = \pm 15$ meV) were taken in a $22.3 \times 22.3$ nm$^2$ area. $g(r, E = \pm 14$ meV) were taken in a $28.2 \times 28.2$ nm$^2$ area. $g(r, E = \pm 12$ meV) were taken in a $20.7 \times 20.7$ nm$^2$ area. Other $g(r, E)$ were taken in $24 \times 24$ nm$^2$ areas. Because the resolution of FT-QPI patterns $|\delta q| = a_0/L \times (2\pi/a_0)$ is dependent on the length ($L$) of FOV, different $|g(q, E)|$ has different $q$-resolution. By selecting the proper number of pixels $N$, the side length of all $|g(q, E)|$ is chosen to be $N|\delta q| = 2.95 \times (2\pi/a_0)$.

## IV. Relationship between Orbital Characters and Scattering Patterns
### A. $q_3$ and $q_2$ Scattering Patterns

It is widely accepted that the staggered arrangement of Se/Te atoms doubles the primitive unit cell, folding the electron pockets on top of each other. However, an underappreciated fact is that the pockets do not become identical replicas and the folded bands appear to have weaker spectral weight in the folded BZ[14] (gray dashed ellipses in Figure 2b). As a result, the $q_1$, $q_2$ and $q_3$ scattering patterns do not have identical geometries, though they only differ by reciprocal vectors in the folded BZ.

Here, we adopt the 1-Fe unit cell, which is more straightforward to analyze $q_3$ and $q_2$ scattering patterns. Accordingly, the *x* and *y* axes used for labeling *d* orbitals refer to the directions along nearest-neighbor Fe atoms. On the single elliptic electron pocket, the segments in the pole regions are dominated by the $d_{xy}$ orbitals, while the flat parts chiefly derive from the $d_{xz}/d_{yz}$ orbitals (Fig. S8a)[17]. The $q_3$ and $q_2$ patterns in our 1-UC Fe(Se,Te) (Fig. S8d) have exactly the same structures as that of previous works on 1-UC FeSe[12,14] and bulk (Li$_{1-x}$Fe$_x$)OHFeSe[6,18,19]. This universality arises from the same geometry and orbital character of the FSs in the 1-UC FeSe-based system.

Specifically, $q_3$ pattern, despite only differing from $q_1$ pattern by a reciprocal vector in the 1-Fe BZ, shows an incomplete elliptic shape with long axis portion missing (Fig. S8d). Previous research has shown that $q_3$ pattern at $(0, \pm 2\pi)$ and $(\pm 2\pi, 0)$ are mostly contributed by the intra-orbital scattering of $d_{xz}$ orbital ($d_{xz} \leftrightarrow d_{xz}$) and $d_{yz}$ orbital ($d_{yz} \leftrightarrow d_{yz}$), respectively[14]. Whereas, $q_2$ pattern, as far as we know, mainly consists of the intra-orbital scattering of $d_{xy}$ orbital ($d_{xy} \leftrightarrow d_{xy}$). The structure of $q_2$ pattern comprises four arcs[12,19], but three of them usually connect jointly, only leaving the arc with smaller scattering vectors separately (Fig. S8d). There are three possible components in $q_2$ pattern, including intra-orbital scattering of $d_{xy} \leftrightarrow d_{xy}$, intra-orbital scattering of $d_{xz} \leftrightarrow d_{xz}$ and $d_{yz} \leftrightarrow d_{yz}$, and inter-orbital scattering of $d_{xz} \leftrightarrow d_{yz}$, $d_{xy} \leftrightarrow d_{xz}$ and $d_{xy} \leftrightarrow d_{yz}$. The latter two components apparently have lower intensity compared to the first one. For one thing, $d_{xz} \leftrightarrow d_{xz}$ and $d_{yz} \leftrightarrow d_{yz}$ scatterings occur between the pole regions and the flat parts of the electron pockets. However, the composition of $d_{xz}/d_{yz}$ orbitals is very limited at the pole regions of the electron pockets[17], and accordingly the joint DOSs of this scattering are relatively small. Furthermore, the center of the arcs formed by $d_{xz} \leftrightarrow d_{xz}$ and $d_{yz} \leftrightarrow d_{yz}$ scattering should be along the directions at $\theta = 0° + 90° \times N$, $N = 0, 1, 2, 3$ (in 1-Fe BZ), which is contradictory to the arcs in $q_2$ pattern centering along the directions at $\theta = 45° + 90° \times N$ (Fig. S8d). For the other, inter-orbital scatterings of $d_{xz} \leftrightarrow d_{yz}$, $d_{xy} \leftrightarrow d_{xz}$ and $d_{xy} \leftrightarrow d_{yz}$ are barely concerned in iron-based superconductors and have negligible effect on the QPI[16,20,21]. Herein, we also do not consider inter-orbital scatterings. Then, only intra-orbital scattering of $d_{xy} \leftrightarrow d_{xy}$ in principle can significantly contribute to the intensity in $q_2$ pattern. More importantly, considering the *k*-space structure of $d_{xy}$ orbital, $d_{xy} \leftrightarrow d_{xy}$ scattering do form the arc structure centering along the directions at $\theta = 45° + 90° \times N$. One may argue that the absence of $d_{xy} \leftrightarrow d_{xy}$ scattering in $q_3$ pattern indicates the suppression of $d_{xy}$ signals in STM measurements[14,22]. Note that $q_3$ pattern at $(0, \pm 2\pi)$ is different from the $q_3$ pattern at $(\pm 2\pi, 0)$ according to a finite Wannier function width[14]. Thus, the particular Wannier function of $d_{xy}$ orbital might selectively inhibit $d_{xy} \leftrightarrow d_{xy}$ scattering in $q_3$ pattern.

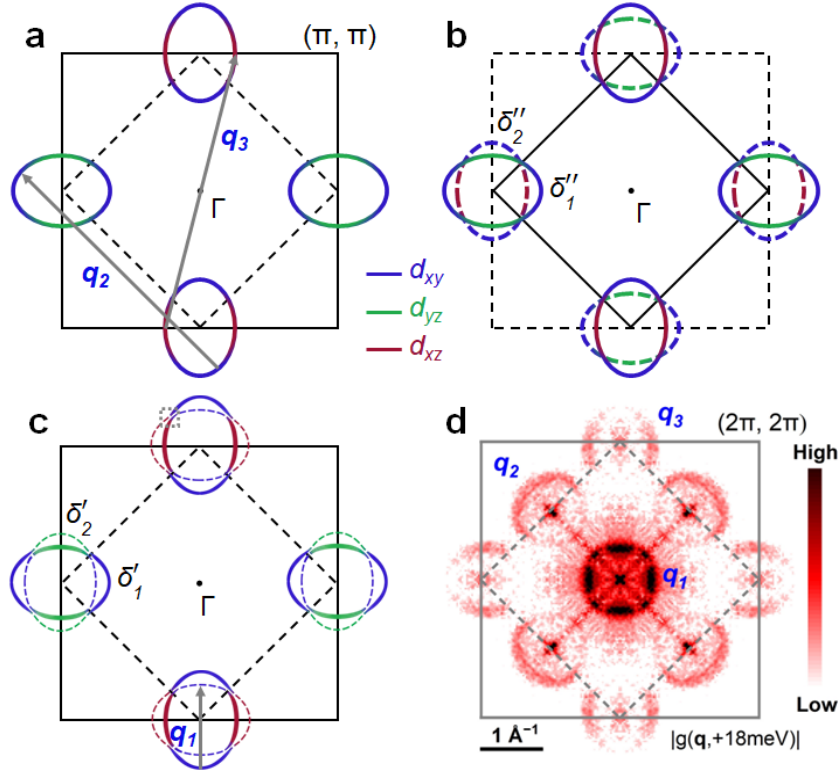

**Figure S8. Orbital characters of electron pockets and relationship between Fe 3$d$ orbitals and scattering patterns. a**, FSs together with orbital characters in 1-Fe BZ (black solid lines). The gray solid arrows depict the representative scattering vectors corresponding to the scattering patterns in **d**. The orbital characters are shown in differernt colors. **b**, FSs in 2-Fe BZ (black solid lines). The dashed ellipses represent the folded band $\delta_2''$, which have the same line width as the original band $\delta_1''$ (solid ellipses) to indicate the same spectral weight. **c**, FSs in unfolded BZ, obtained by unfolding the 2-Fe unit cell band structures, which is different from the original 1-Fe BZ since it incorporates the translational symmetry breaking potential arising from the staggered arrangement of Se/Te atoms[24]. The dashed ellipses represent the unfolded ("shadow") band $\delta_2'$, which have smaller line width than the original band $\delta_1'$ (solid ellipses) to imply the weaker spectral weight. The small gray dashed square highlights the gap due to the band hybridization. The gray solid arrow depicts the inter-pocket scattering in $q_1$ pattern. **d**, FT-QPI pattern of $|g(q, E = 18\text{ meV})|$, which rotates 45° compared to Figure 4a, in order to correspond to the direction of 1-Fe BZ.

## B. $q_1$ Scattering Pattern

While $q_3$ and $q_2$ scattering patterns can be well explained in 1-Fe BZ, the real symmetry of the 1-UC FeSe-based system actually dictates 2-Fe unit cell. It is widely accepted that, due to the breaking of translational symmetry, originating from the staggered arrangement of Se/Te atoms, bands in the 1st BZ of 1-Fe unit cell get 'folded' into the 1st BZ of 2-Fe unit cell (Fig. S8b). However, an underappreciated fact is that, regarding the weak symmetry breaking, experimental studies such as ARPES and STM normally show different band structures in *different* BZ of the 2-Fe unit cell rather than displaying all the bands in the 1st BZ of the 2-Fe unit cell[23]. Therefore, for a direct comparison with experimental results, it is necessary to theoretically unfold the 2-Fe unit cell band structures. The unfolded band structures (Fig. S8c) are different from the nominal 1-Fe unit cell band structure (Fig. S8a), since it considers the translational symmetry breaking potential, which generates a "shadow" band $\delta_2'$ in addition to the original band $\delta_1'$. More importantly, the "shadow" band $\delta_2'$ in unfolded BZ is also different from the folded band $\delta_2''$ in 2-Fe BZ (Fig. S8b)[24]. For one thing, $\delta_2'$ possesses overall weak spectral weight, consistent with the experimental observation[9], whereas the $\delta_2''$ should have almost same spectral weight with the original band from the perspective of theoretical view. For another, the orbitals switch parity upon band unfolding, where the parity is defined as the mirror symmetry with respect to the Fe plane. For example, the odd parity orbitals of $d_{xz}$ (red) and $d_{yz}$ (green) in $\delta_2''$ change to even parity orbital of $d_{xy}$ (blue) in $\delta_2'$, and vice versa. The orbital-parity switching is dictated by the glide-mirror symmetry, preserving by the staggered arrangement of Se/Te atoms[14,24].

When investigating the $q_1$ pattern, it is prerequisite to incorporate either the "shadow" band $\delta_2'$ (from experimental perspective) or the folded band $\delta_2''$ (from theoretical perspective). Then, the band hybridization would open a gap at the intersection of the two elliptic-shaped electron pockets (highlight by the small gray dashed square in Fig. S8c). After that, the inner and outer electron pockets ($\delta_1$ and $\delta_2$ in Figure 2b) are formed, where the pairing structure and the pairing symmetry such as 'bonding-antibonding' $s_\pm$-wave or 'quasi-nodeless' d-wave can be defined. It should be noted that the unfolded BZ has the superiority in demonstrating the inter-pocket scattering in $q_1$ pattern (gray solid arrow in Fig. S8c), since it is intra-orbital scattering. In contrast, the inter-pocket scattering in $q_1$ pattern is inter-orbital scattering $d_{xy}\leftrightarrow d_{xz}/d_{yz}$ in 2-Fe BZ (Fig. S8b), which should be suppressed[20].

The last thing deserving clarification is that, despite the weak spectral weight of "shadow" band, the intensity of inter-pocket scattering is still noticeable and even dominate over the $q_1$ pattern, which mainly benefits from the more coherent sign-changing scattering channel. In general, the intensity of FT-QPI pattern is determined by not only the DOSs, but also the Bogoliubov coherence factors, given by[25]

$$v_i(\boldsymbol{k}) = \text{sign}[\Delta_i(\boldsymbol{k})]\sqrt{\frac{1}{2}\left[1 - \frac{\varepsilon_i(\boldsymbol{k})}{E_i(\boldsymbol{k})}\right]}, \ u_i(\boldsymbol{k}) = \sqrt{\frac{1}{2}\left[1 + \frac{\varepsilon_i(\boldsymbol{k})}{E_i(\boldsymbol{k})}\right]} \quad (6)$$

where $E_{\boldsymbol{k}} = \sqrt{\Delta_{\boldsymbol{k}}^2 + \varepsilon_{\boldsymbol{k}}^2}$ is the Bogoliubov spectrum of superconductors and $i/j$ represents band. The scattering probability for a transition from the initial state ($i,\boldsymbol{k}$) to the final state ($j,\boldsymbol{k'}$) is given by

$$W_{i\to j}(\boldsymbol{k},\boldsymbol{k'}) \propto \left|u_i(\boldsymbol{k})u_j^*(\boldsymbol{k'}) \pm v_i(\boldsymbol{k})v_j^*(\boldsymbol{k'})\right|^2 \cdot |V(\boldsymbol{k'}-\boldsymbol{k})|^2 N_i(\boldsymbol{k})N_j(\boldsymbol{k'}) \quad (7)$$

where $V$ is scattering potential, $N$ is DOSs, and the term $C(\boldsymbol{k},\boldsymbol{k'}) = \left|u_i(\boldsymbol{k})u_j^*(\boldsymbol{k'}) \pm v_i(\boldsymbol{k})v_j^*(\boldsymbol{k'})\right|^2$ derives from the Fermi's golden rule and chooses minus for scalar potential[25,26]. Since the scatters in our system are nonmagnetic (Supporting Information II.A), the value of $C(\boldsymbol{k},\boldsymbol{k'})$ for the sign-changing scattering would be much larger than the one for the sign-preserving scatterings within the energy range of SC gaps ($\varepsilon_{\boldsymbol{k}} \sim 0$). Correspondingly, the inter-pocket sign-changing scattering should have larger intensity than the intra-pocket sign-preserving scattering in $q_1$ pattern, shown as the evident negative signals in PR-QPI patterns (Figure 4 and Fig. S12).

## V. Two Approaches to Extract Gap Structure
### A. Determination of the Contour and Intensity of $q_1$ Pattern

In order to more precisely locate the profile of $q_1$ pattern (Fig. S9a) and extract the intensity correspondingly, we take several line cuts of $|g(\boldsymbol{q}, E)|$ along different radial directions ($\theta$), with step size $\delta\theta = 3°$ and azimuthal averaging over $\theta \pm 1°$. Figure S9b shows typical line cuts of Gaussian-core subtracted FT-QPI intensity. The peaks with sizable broadening, can be well-fitted by Gaussian plus constant background. Sometimes, there are a few peaks close to each other, like the line cut at $\theta = 6°$. In this case, we choose the peak according to the empirical geometry of FSs[17]. Figure S9c displays line cuts without Gaussian core subtraction, where there is an obvious slope. Correspondingly, we performed Gaussian fitting with linear background and obtained almost the same location of the Gaussian peak. The variation between the two approaches is within 0.02 $\pi/a_0$, smaller than the data resolution $|\delta q| = 0.034\ \pi/a_0$ in most FT-QPI patterns (0.063 $\pi/a_0$ in $|g(q, E = 16\ \text{meV})|$). Therefore, the Gaussian core subtraction, despite slightly decreasing the absolute intensity of $q_1$ pattern, does not modify the scattering wave vector.

Herein, we used the Gaussian-core subtracted FT-QPI intensity to obtain Figure 3c in order to better distinguish scattering vectors with short length. The Gaussian core subtraction is unified as $|g_{new}(q, E)| = |g_{raw}(q, E)|\ [1 - 0.85 * Gaussian\ (q = (0, 0), \sigma = 0.18\ \pi/a_0)]$. Then we took line cuts along radial direction θ = 45°, averaged by the nearest-neighbor pixel, and normalized them individually by their averaged intensity within [0.2, 0.8] $\pi/a_0$.

We used the raw FT-QPI intensity to obtain raw intensity $I(\theta, E)$ and further implemented two steps to obtain the normalized intensity $I_n(\theta, E)$ shown in Figure 2f (also Fig. S11c). The first step is to individually divide the raw intensity by its averaged intensity $I_{new}(\theta, E) = I_{raw}(\theta, E)/I_b(E)$. The averaged intensity is calculated within the area of the concentric circles $|q^c_{min}(E)| < |q(E)| < |q^c_{max}(E)|$, where $|q^c_{min}(E)|$ and $|q^c_{max}(E)|$ are the minimum and maximum modulus of scattering vectors on the extracted contour $q^c(\theta, E)$, respectively. This procedure aims to eliminate the influence of different scattering potential, deriving from the varying number and position of defects in QPI measurement at different FOVs. Using the $\hat{T}$-matrix approach[5], the intensity of FT-QPI pattern is given by

$$g(\boldsymbol{q}, \omega) = -\frac{1}{\pi} \sum_{i,j} \text{Tr}\ \text{Im} \int \frac{d^2k}{(2\pi)^2} (\tau_0 + \tau_3) \hat{G}^0_i(\boldsymbol{k}, \omega)\ \hat{t}(\omega)\ \hat{G}^0_j(\boldsymbol{k} + \boldsymbol{q}, \omega) \qquad (8)$$

where $\tau_0$, $\tau_3$ is the Pauli matrix, and $\hat{G}^0_{i/j}(\boldsymbol{k}, \omega)$ represents the Green's function for the band $i/j$. Here we use the general approximation that the $\hat{T}(\omega)$ matrix is momentum independent, which is a characteristic of point like *s*-wave scatterers, given by

$$\hat{T}(\omega) = \sum_j e^{i\boldsymbol{q}\cdot\boldsymbol{R}_j} \left[1 - V_j \int \frac{d^2k}{(2\pi)^2} \hat{G}^0(\mathbf{k}, \omega)\right]^{-1} V_j \qquad (9)$$

The summation is for all defects in the FOV. $\boldsymbol{R}_j$ is the position of the defect $j$ and $V_j$ represents the matrix of its scattering potential. It is obvious that $\hat{T}(\omega)$ changes as the number and position of defects vary, consequently altering the intensity of the scattering pattern.

Though the procedure mentioned above can eliminate the variation of the $\hat{T}(\omega)$, it also wipes out the dependence of FT-QPI intensity on energy $\omega$, originating from $\hat{G}^0(\mathbf{k}, \omega)$. Hence, the second step is to use the DOSs, characterized by the averaged tunneling spectrum in Fig. S1b, to restore the dependence on $\omega$, via further multiplying the intensity by the square of the corresponding DOSs $I_n(\theta, E) = I_{new}(\theta, E) * [dI/dV(E)]^2$. The reason for this is that scattering intensity positively correlates with both the DOSs of initial state and final state.

After the normalization procedure, the intensity is virtually equivalent to

$$g'(\mathbf{q}, \omega) = \frac{\frac{1}{2}\text{Tr}\ \text{Im} \sum_{\mathbf{k}} (\tau_0+\tau_3)\hat{G}^0(\mathbf{k},\omega)\hat{T}(\omega)\hat{G}^0(\mathbf{k+q},\omega)}{\sum_{\mathbf{q}} \frac{1}{2}\text{Tr}\ \text{Im} \sum_{\mathbf{k}} (\tau_0+\tau_3)\hat{G}^0(\mathbf{k},\omega)\hat{T}(\omega)\hat{G}^0(\mathbf{k+q},\omega)} \cdot (\frac{1}{2}\text{Tr}\ \text{Im} \sum_{\mathbf{k}} (\tau_0 + \tau_3)\hat{G}^0(\mathbf{k}, \omega))^2 \qquad (10)$$

Therefore, we unified the FT-QPI patterns measured at different FOVs and simultaneously, albeit not strictly, recovered their energy dependence, making the intensity ready for direct comparison.

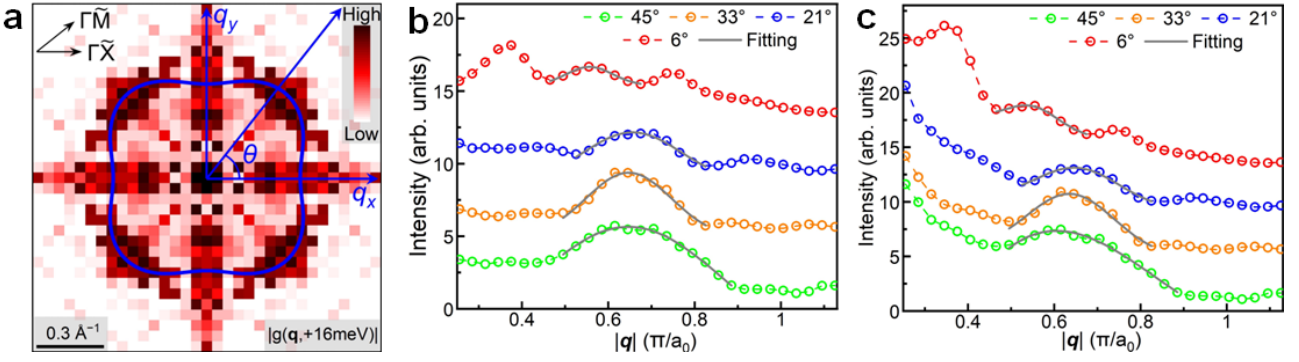

**Figure S9. Extracting the $q_1$ pattern. a**, $q_1$ pattern of $|g(q, E = 16 \text{ meV})|$, where $\theta$ is defined relative to the $q_x$ axis (corresponding to the $\Gamma\tilde{X}$ direction in 2-Fe BZ). The blue closed curve fitted by $[R - r * \cos 4\theta]$ more accurately sketches the profile of $q_1$ pattern. **b**, The line cuts of Gaussian-core subtracted FT-QPI intensity versus $q$ along different radial direction $\theta = 6°, 21°, 33°$ and $45°$ (vertically offset for clarity). The gray solid lines are the Gaussian fits. The obtained positions of the Gaussian peak are 0.556 $\pi/a_0$ (6°), 0.656 $\pi/a_0$ (21°), 0.647 $\pi/a_0$ (33°) and 0.647 $\pi/a_0$ (45°). **c**, Same line cuts as **b** but for FT-QPI intensity without Gaussian core subtraction. The gray solid lines are the Gaussian fits plus linear background. The intensities $I(\theta, E)$ of the Gaussian peak are further processed to give $I_n(\theta, E)$ in Fig. S11c. The obtained positions of the Gaussian peak are 0.552 $\pi/a_0$ (6°), 0.667 $\pi/a_0$ (21°), 0.652 $\pi/a_0$ (33°) and 0.668 $\pi/a_0$ (45°).

## B. Determination of FS from Maximum-Intensity Points in the |q|-E plane

Normally, the tips of the constant-energy-contours (CECs) lie on the same contour as the FS of the underlying normal state band structure, since the SC gap opens particle-hole symmetric relative to the FS[16]. Then the determination of FS is purely a mathematical process. See Fig. S10 and Equation (11,12).

$$x_1 = \frac{\sqrt{2}}{4}\left(|q_M^{t_1}| + |q_M^{t_2}|\right) \tag{11}$$

$$y_1 = \frac{\sqrt{2}}{4}\left(|q_M^{t_1}| - |q_M^2|\right) \tag{12}$$

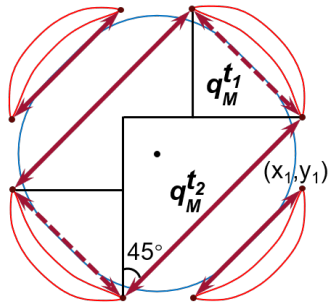

**Figure S10. Sketch to obtain Fermi surface.** Red contours show CECs of $\delta_2$. Red solid arrows depict the scatterings vectors associated with 'banana' tips along $\theta = 45°$. Note that red dashed arrows have the same length as the shorter red solid arrows. The black lines are deliberately plotted to help understand the Equation (11,12). The blue contour represents FS of $\delta_1$.

## C. Determination of Gap structure and FS from the Iso-intensity Line in the θ-E plane

Another approach to distill gap structure also utilizes the scattering vectors associated with 'banana'

tips. The difference lies in the selection of these vectors. While the approach in main body is based on the scattering along a specific high-symmetry direction (Figure 3a), which only change length as energy varies, this approach focuses on scatterings at opposite side ($\boldsymbol{k} \leftrightarrow \boldsymbol{-k}$), which change direction as energy varies, shown as red arrows in Fig. S11a. Fig. S11b shows the schematic scattering intensity, deriving from the scattering between CECs at opposite side, where the maximum intensity points correspond to the scatterings between tips of CECs. Outside the CECs, ideally, there is no spectral weight, and the intensity shall reduce immediately and dramatically. Essentially, intensity $I(\theta, E)$ in Fig. S11b is a projection of the FT-QPI intensity $|g(\boldsymbol{q}, E)|$ onto the band dispersion $E(\boldsymbol{q}) = E(2\boldsymbol{k})$, and concurrently maps the coordinate from two-dimensional $\boldsymbol{q}$ to one-dimensional $\theta$. In this $\theta$-$E$ plane, the maximum-intensity points can be expressed as $\theta_M = 45° \pm \theta_0 + 90° \times N$ (Fig. S11b). As $E$ increases from $|\Delta_2^{min}|$ to $|\Delta_2^{max}|$, $\theta_0$ increases from 0° to 45°, and this curved trajectory contains the information about the gap structure.

In Supporting Information V.A, we extract the contour of $\boldsymbol{q_1}$ scattering pattern. Then, the normalized intensity on $\boldsymbol{q_1}$ scattering pattern at different $E$ is shown in Fig. S11c. Contrary to expectations, the maximum-intensity points are inconspicuous (or unreasonable) in Fig. S11c. We ascribe the absence of maximum-intensity points in experiments to the multiband effect. Recalling that the three sub-sets of $\boldsymbol{q_1}$ scattering pattern cannot be distinctively resolved in the QPI measurement, the $\boldsymbol{q_1}$ contour extracted above is more likely controlled by the inter-pocket scattering of $\boldsymbol{\delta_1} \leftrightarrow \boldsymbol{\delta_2}$ rather than the desired intra-pocket scattering of $\boldsymbol{\delta_2} \leftrightarrow \boldsymbol{\delta_2}$. This is also evidenced by the PR-QPI measurement (blue pixels in Figure 4) and has substantial theoretical basis (see Supporting Information IV.B for the discussion on Bogoliubov coherence factors). In addition, intra-pocket scattering of $\boldsymbol{\delta_1} \leftrightarrow \boldsymbol{\delta_1}$ also contributes to the intensity on $\boldsymbol{q_1}$ scattering pattern, especially around the $\Gamma\tilde{X}$ direction, where the three sub-sets merge. This further diminishes the distinction of the maximum-intensity points.

Although the maximum-intensity points are obscure and hard to precisely discern, there is still possible to roughly approximate the gap structure[18]. Note that the black dashed curve in Fig. S11b depicts an iso-intensity line, which has a similar line shape as the curve connecting maximum intensity points. The intuition that FT-QPI intensity is suppressed along gap maximum direction (when $E$ < $\Delta_{max}$), and the location where FT-QPI intensity first emerges would be associated with gap minimum direction, is widely accepted[16,27]. The iso-intensity line approximation is essentially a simplified way to make this intuition quantitative. Regarding the selection of the iso-intensity line, we chose the intensity $I_0$ to match the maximum of the anisotropic gap $|\Delta_2^{max}| = 18$ meV, which can be recognized from the average tunneling spectrum (Fig. S1d). Then the nominal iso-intensity line (white curve in Fig. S11c) was acquired via tracking the trajectory of energy thresholds (black hollow dots in Fig. S11c). The energy thresholds were extracted when the intensity, along $E$-axis direction at a fixed $\theta$, starts to be larger than $I_0$. One can see that the iso-intensity line clearly outlines the intensity profile associated with the evolution of the CECs and yields the gap minimum $|\Delta_2^{min}| = 11.6$ meV (Fig. S11c). We show the approximated gap structure and FS, obtained by iso-intensity line approach, in Fig. S11d. The anisotropic gap structure is consistent with the result in Figure 3d, although the shapes of FS show some difference. The FS obtained by iso-intensity approach is almost circle-shape (Fig. S11d), which looks quite similar to the inter-pocket scattering pattern in theoretical simulations[6,12]. This once again indicates the intensity in Fig. S11c comes from the inter-pocket scattering. Note that the iso-intensity approach is merely an approximation of gap structure, since the FT-QPI intensity not only is related to the gap structure, but may also be influenced by the variations in the orbital character[17,20].

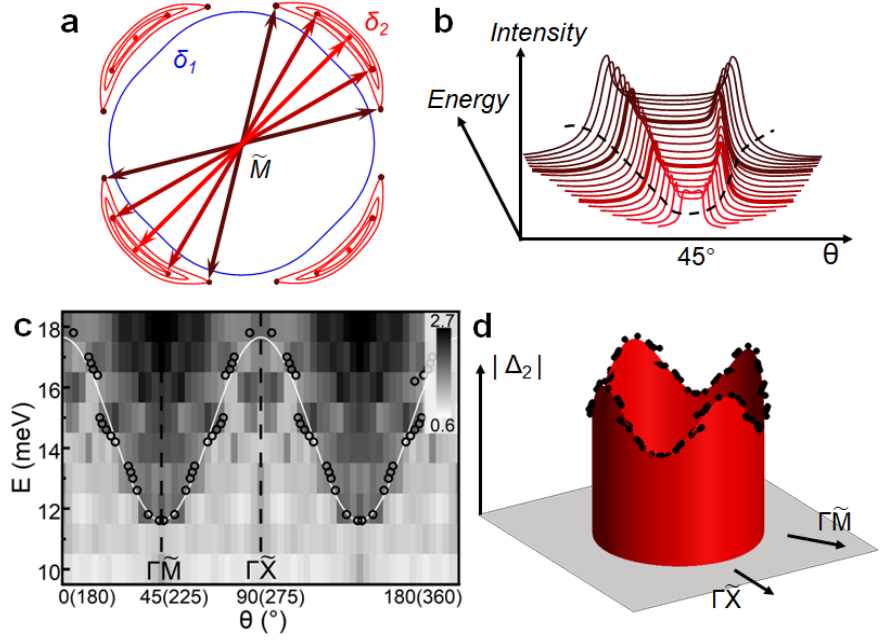

**Figure S11. Determination of gap structure by iso-intensity line approach. a**, The same schematic CECs as Figure 3a. But in this case, the arrows associated with 'banana' tips change direction as the size of CECs enlarges. **b**, Schematic scattering intensity between CECs ($\delta_2 \leftrightarrow \delta_2$), where gradually darkened color indicates scattering intensity at increasing energy. The maxima correspond to the scatterings between the tips of CECs. The black dashed curve depicts the iso-intensity line, which has a similar line shape as the curve connecting maximum intensity points. **c**, Two-dimensional map of normalized $I_n(\theta, E)$. The black hollow dots locate the energy thresholds where the intensity along $E$-axis at a fixed $\theta$ start to be larger than $I_0 = 1.475$. The white curve, which is the nominal iso-intensity line, schematically depicts the trajectory of these energy thresholds. **d**, The approximation for gap structure (black solid dots) on the simultaneously obtained FS of $\delta_2$, following the iso-intensity approach in **c**.

## VI. Phase-referenced (PR-) QPI Approach
### A. Introduction to different PR-QPI approaches

Early PR-QPI approaches developed for cuprates encounter difficulties when applied to FeBSs. For instance, QPI with vortex cores as scattering centers in principle can offer phase sensitive information, since vortices as magnetic scattering potentials can enhance the signal of certain scattering wave vectors[28]. However, vortices are spatially more extended in FeBSs, which complicates comparison with theoretical calculations that assume point-like scattering potentials, making the interpretation of field-dependent QPI experiments disputed. Hereafter, new approaches to visualize the symmetry of the SC gaps in FeBSs have been proposed. Among them, the approach by Hirschfeld, Altenfeld, Eremin, and Mazin (HAEM) is based on an integral over the real part of anti-symmetrized inter-pocket FT-QPI scattering pattern[29]. The original HAEM approach requires one single impurity sited at the exact center in QPI mappings[6,15]. Nevertheless, in order to obtain a high $q$ resolution, we measured QPI mappings in large FOVs, which inevitably contains multiple native defects. And it is difficult for original HAEM approach to eliminate the irrelevant phase information coming from the position of defects, although there have been investigations to do so[30]. Another approach, called defect-induced bound states QPI (DBS-QPI), extracts the scattering phase of FT-QPI at positive energy $\theta_{q,E}$ and negative energy $\theta_{q,-E}$, and using the phase difference $\theta_{q,-E} - \theta_{q,E}$ to offset phase coming from irrelevant details of defects[31,32]. Thus, the obtained DBS-QPI can clearly display phase information associated with SC gaps[33]. However, the DBS-QPI approach is limited to the energies of defect-induced bound states. Even though later experiments used DBS-QPI approach at the energies near the SC gap edge or within SC gap, these researches still rely on defect-induced bound states, which are supposed to appear near the gap edges and mix with the gap peaks[34] or produce a continuous change of density of states within the SC gap rather than generate sharp bound state peaks[35].

Here, we adopted the data process in DBS-QPI approach, which defines the signals at positive and negative energies as[31]

$$g_{pr}(\mathbf{q}, +E) = |g(\mathbf{q}, +E)|\cos\left(\theta_{\mathbf{q},+E} - \theta_{\mathbf{q},-E}\right) \tag{13}$$

$$g_{pr}(\mathbf{q}, -E) = |g(\mathbf{q}, -E)| \tag{14}$$

However, our PR-QPI approach is based on the energy range between the gap maximum $|\Delta_2^{max}|$ and the gap minimum $|\Delta_2^{min}|$, which is different from the DBS-QPI approach relying on the energy of defect bound states. Essentially, our PR-QPI approach originates from the HAEM approach, albeit more simplified (see theoretical analysis in Supporting Information VI.B).

When there are numerous defects in the FOV, the measured differential conductance mapping $g(r, \pm E)$ can be viewed as the superposition of the standing waves from multiple scattering centers, given by

$$g(\mathbf{r}, \pm E) = \sum_j g^s(\mathbf{r} - \mathbf{R}_j, \pm E) \tag{15}$$

After the FT, the results are

$$g(\mathbf{q}, +E) = g^s(\mathbf{q}, +E) \sum_j e^{-i\mathbf{q}\cdot\mathbf{R}_j} \tag{16}$$

$$g_{pr}(\mathbf{q}, +E) = \left|g^s(\mathbf{q}, +E) \sum_j e^{-i\mathbf{q}\cdot\mathbf{R}_j}\right| \cos(\theta^s_{\mathbf{q},+E} - \theta^s_{\mathbf{q},-E}) \tag{17}$$

Therefore, the data process can offset the disturbing phase contribution stemming from the random distribution of defects and accurately reflect the phase information associated with SC gaps[31]. However, this universality comes at a cost; it will bring a random amplitude factor ($|\sum_j e^{-i\mathbf{q}\cdot\mathbf{R}_j}|$) to PR-QPI signal, which recommends integrating $g_{pr}(\mathbf{q}, +E)$ over the inter-pocket scattering region to obtain the actual intensity[32]. Herein, we mainly concentrated on the phase information, i.e. the sign of the PR-QPI signals, regardless of the fluctuation of the signal amplitude. Moreover, recalling the anisotropic structure of the SC gap, the integral at the energy within $|\Delta_2^{min}|$ and $|\Delta_2^{max}|$ fails in principle. Hence, we analyzed our PR-QPI data without defining an integral quantity.

## B. $\hat{T}$-matrix approach in Born approximation

We summarize the main results of our PR-QPI response within the $\hat{T}$-matrix approach in Born approximation[5]. Out of convenience, we consider the case that one single non-magnetic defect sited at the center in the FOV, which can be extended to the multiple defects case as mentioned in Supporting Information VI.A. Theoretically, the zero-temperature intensity of the FT-QPI pattern is given by Equation (8)[20]. The Born approximation for the multiple scattering by a single impurity in the band and Nambu-Gorkov space is defined as

$$\hat{t}(\omega) = V\tau_3 + V\tau_3 \int \frac{d^2k}{(2\pi)^2} \hat{G}^0(\boldsymbol{k}, \omega) V\tau_3 + \left[V\tau_3 \int \frac{d^2k}{(2\pi)^2} \hat{G}^0(\boldsymbol{k}, \omega)\right]^2 V\tau_3 + \cdots$$
$$= \left[I - V\tau_3 \int \frac{d^2k}{(2\pi)^2} \hat{G}^0(\boldsymbol{k}, \omega)\right]^{-1} V\tau_3 \approx V\tau_3 \tag{18}$$

where higher-order scatterings can be negligible, when the scattering potential is weak. The Nambu-Gorkov Green's function for band $i$ is given by

$$\hat{G}_i^0(\boldsymbol{k}, \omega) = \frac{1}{\omega + i\eta - B_i(\boldsymbol{k})} \tag{19}$$

where $\eta$ expresses the energy broadening and $B_i(\boldsymbol{k})$ is defined as

$$B_i(\boldsymbol{k}) = \begin{bmatrix} \varepsilon_i(\boldsymbol{k}) & \Delta_i(\boldsymbol{k}) \\ \Delta_i(\boldsymbol{k}) & -\varepsilon_i(\boldsymbol{k}) \end{bmatrix} \tag{20}$$

Regarding the ring-like structure of FT-QPI patterns, the intensities are contributed by the scatterings between $\boldsymbol{k}$ and $\boldsymbol{k}'$ located on the CECs, especially where the tangents of the CECs are parallel.

$$\boldsymbol{q} = \boldsymbol{k}' - \boldsymbol{k}, \quad \varepsilon_i(\boldsymbol{k}) = \sqrt{\omega^2 - \Delta_i^2(\boldsymbol{k})}, \quad \varepsilon_j(\boldsymbol{k}') = \sqrt{\omega^2 - \Delta_j^2(\boldsymbol{k}')} \tag{21}$$

Therefore, we only consider the parallel-tangent scattering hereinafter, which as far as we know dominates the scattering intensity. Since the intensity of a specific scattering vector $\boldsymbol{q}$ now solely corresponds to the scattering between two local momenta, the Equation (8) is greatly simplified as

$$g(\boldsymbol{q}, \omega) = -\frac{1}{\pi} \sum_{i,j} \mathrm{Tr}\, \mathrm{Im}(\tau_0 + \tau_3) \hat{G}_i^0(\boldsymbol{k}, \omega) \hat{t}(\omega) \hat{G}_j^0(\boldsymbol{k}', \omega) \tag{22}$$

Moreover, this simplification also refines the effect of the anisotropic SC gap, since the FT-QPI signal of a specific scattering vector $\boldsymbol{q}$ now only relates to the fixed values of $\Delta_i(\boldsymbol{k}) = \Delta_i$ and $\Delta_j(\boldsymbol{k}') = \Delta_j$. Then the Equation (22) can be calculated as

$$g(\boldsymbol{q}, \omega) \approx -0.02 V (\hbar v)^{-\frac{3}{2}} \kappa^{\frac{1}{2}} (\omega^2 - \Delta_i^2)^{-\frac{3}{8}} (\omega^2 - \Delta_j^2)^{-\frac{3}{8}} |\omega|^{-\frac{1}{2}} \eta^{-\frac{1}{2}}$$
$$\cdot \left[\left(\omega + \sqrt{\omega^2 - \Delta_i^2}\right)\left(\omega + \sqrt{\omega^2 - \Delta_j^2}\right) - \Delta_i \Delta_j\right] \mathrm{sgn}(\omega) \tag{23}$$

where $\hbar v = d\varepsilon/dk$ is band dispersion and $\kappa = (d^2k/d\theta^2)/(dk/d\theta)^2$ is the curvature of CECs on Fermi level, both averaged over $i, j$. Besides, it requires $|\omega| > \max[|\Delta_i(\boldsymbol{k})|, |\Delta_j(\boldsymbol{k}')|]$ to promise the non-zero DOSs on $\boldsymbol{k}$ and $\boldsymbol{k}'$. The most important conclusion from Equation (23) is

$$\Delta_i \Delta_j < 0 \Leftrightarrow g(\boldsymbol{q}, \omega) g(\boldsymbol{q}, -\omega) < 0 \tag{24}$$

This is exactly the principle of our PR-QPI approach. The opposite sign indicates a nearly $\pi$ phase shift between FT-QPI signals on the positive and negative energy sides, which yields $g_{pr}(\boldsymbol{q}, +E) < 0$.

Accordingly, the results of intra-pocket sign-preserving scattering are given by

$$g(\boldsymbol{q}, \omega) \approx -0.04 V (\hbar v)^{-\frac{3}{2}} \kappa^{\frac{1}{2}} (\omega^2 - \Delta^2)^{-\frac{1}{4}} |\omega|^{-\frac{1}{2}} \eta^{-\frac{1}{2}} \left(\omega + \sqrt{\omega^2 - \Delta^2}\right) \mathrm{sgn}(\omega) \tag{25}$$

$$g(\boldsymbol{q}, \omega) g(\boldsymbol{q}, -\omega) > 0 \tag{26}$$

The intensity in Equation (25) is smaller than the sign-changing scattering intensity in Equation (23), consistent with the conclusion from the joint density of states model, that the value of $C(\boldsymbol{k}, \boldsymbol{k}')$ for the sign-changing scattering would be much larger than the one for the sign-preserving scatterings (Supporting Information IV.B).

In comparison, for magnetic scattering potential, the intensity is given by

$$g(\boldsymbol{q},\omega) \approx -0.02V(\hbar v)^{-\frac{3}{2}}\kappa^{\frac{1}{2}}(\omega^2 - \Delta_i^2)^{-\frac{3}{8}}(\omega^2 - \Delta_j^2)^{-\frac{3}{8}}|\omega|^{-\frac{1}{2}}\eta^{-\frac{1}{2}}$$
$$\cdot \left[\left(\omega + \sqrt{\omega^2 - \Delta_i^2}\right)\left(\omega + \sqrt{\omega^2 - \Delta_j^2}\right) + \Delta_i\Delta_j\right]\text{sgn}(\omega) \quad (27)$$

The conclusion is exactly opposite to the non-magnetic case as
$$\Delta_i\Delta_j < 0 \Leftrightarrow g(\boldsymbol{q},\omega)g(\boldsymbol{q},-\omega) > 0 \quad (28)$$

Another aspect is that whether the aforementioned conclusion changes for strong scattering potential when higher order scatterings need to be considered (beyond Born approximation). If the product of scattering potential and momentum integrated Green function approaches unity, the $\hat{T}$-matrix achieves a resonance and becomes close to 'singularity' (see Equation (18)). An effect of this 'singularity' in $\hat{T}$-matrix comes into play if the imaginary part of the momentum integrated Green function is small, i.e. the density of states is small. This only emerges below the energy of the SC gap minimum, where the density of states changes strongly. Therefore, the presence of 'singularity' could diminish the amplitude of the PR-QPI signals near $|\Delta_2^{min}|$, and yet leaves the PR-QPI signals at higher energies mostly unaffected. Moreover, in our model, the 'singularity' effect is not enough to change the sign of the PR-QPI signals at $|\Delta_2^{min}|$ as long as the scattering potential $V$ is significantly smaller than the energy scale associated with the bandwidth of the normal state electronic structure.

## C. Gap Symmetry revealed by PR-QPI in 1-UC Fe(Se,Te)/STO

Figure S12 shows the PR-$\boldsymbol{q_1}$ patterns. At $E$ = 10 meV, the PR-$\boldsymbol{q_1}$ pattern conspicuously presents positive signals, which is in accordance with the obtained gap minimum $|\Delta_2^{min}| \approx$ 10 meV. As energy increases, the negative signals emerge in $\Gamma\widetilde{M}$ directions and gradually expand to $\Gamma\widetilde{X}$ directions along the contour of $\boldsymbol{q_1}$ pattern, indicating the anisotropic gap structure and the sign reversal gap symmetry in 1-UC Fe(Se,Te). In order to manifest the evolution more quantitatively, we use the same data processing procedure as Figure 3c to extract the line cut of phase part ($\cos(\theta_{\boldsymbol{q},+E} - \theta_{\boldsymbol{q},-E})$) in PR-QPI pattern, and use the line cut to determine the sign of the maximum-intensity points in FT-QPI patterns (colored dots in Figure 3c). The maximum-intensity points at energy within 18 meV and 14 meV have negative phase part in PR-QPI, while they have nearly vanishing phase part at energy within 13 meV and 11 meV, and finally have distinctly positive phase part at $E$ = 10 meV. If the higher order terms in the $\hat{T}$-matrix approach cannot be neglected any more, i.e. approaching the 'singularity' in the $\hat{T}$-matrix, the sign-changing scattering would be weakened, possibly resulting in a reverse of the phase part. We discuss in Supporting Information VI.B the conditions under which the 'singularity' is not strong enough to change the sign of PR-QPI signals. Here, the positive signals at $E$ = 10 meV originate from the intra-pocket scattering of $\boldsymbol{\delta_1} \leftrightarrow \boldsymbol{\delta_1}$ instead of the 'singularity' effect (Figure 4h).

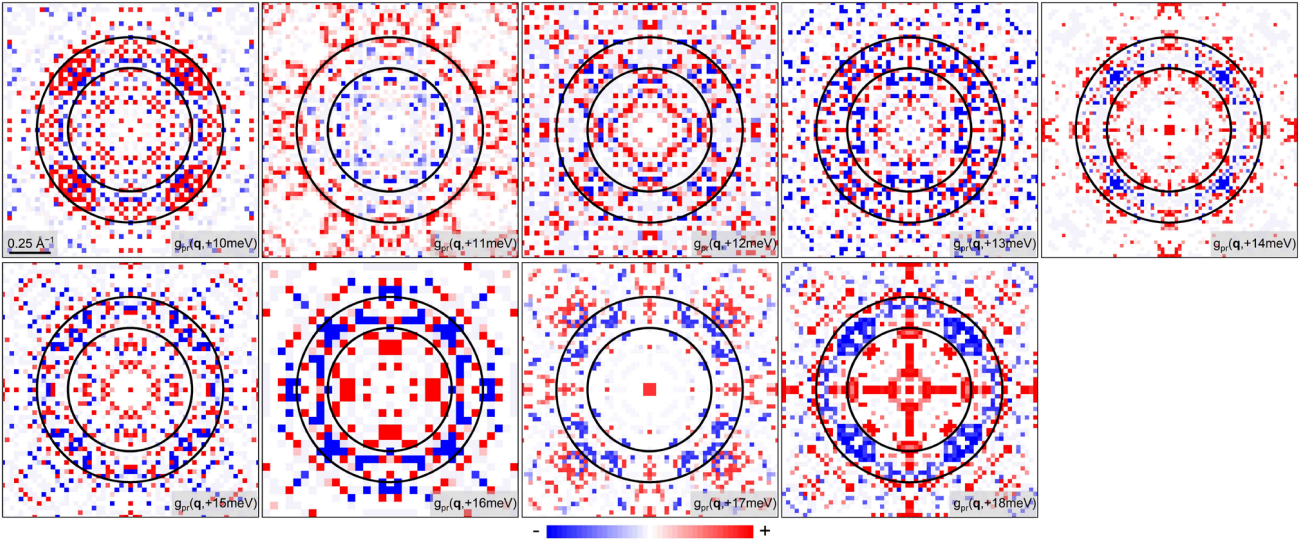

**Figure S12. The central PR-QPI patterns** $g_{pr}(\boldsymbol{q},+E)$. The region within the concentric black circles schematically covers the $\boldsymbol{q}_1$ pattern, with inner radius as 0.5 $\pi/a_0$ and outer radius as 0.75 $\pi/a_0$. The resolution of each $g_{pr}(\boldsymbol{q},+E)$ is same as the resolution of its $|g(\boldsymbol{q},E)|$ (Fig. S7), and we chose the side length of all $g_{pr}(\boldsymbol{q},+E)$ to be 2 $\pi/a_0$.

At the gap maximum ($E$ = 18 meV), the PR-QPI pattern is most clearly resolved (Figure 4e) and show two striking phenomena. For one thing, although the red-blue-red pixels along $\Gamma\widetilde{M}$ directions are consistent with both 'bonding-antibonding' $s_\pm$-wave and 'quasi-nodeless' $d$-wave, the noticeably red pixels in $\Gamma\widetilde{X}$ directions in Figure 4e prefer 'bonding-antibonding' $s_\pm$-wave scenario to 'quasi-nodeless' $d$-wave scenario. For another, the PR-$\boldsymbol{q}_2$ scattering pattern at 18 meV (Figure 4e) shows the outer positive-signal arcs and the middle negative-signal arcs, verifying the sign reversal gaps on the two pockets, once again. However, compared to the $\boldsymbol{q}_1$ ring, the inner positive-signal arcs disappear, which may arise from two aspects. For one thing, while intra-orbital scatterings of $d_{xz}\leftrightarrow d_{xz}$ and $d_{yz}\leftrightarrow d_{yz}$ significantly contribute to the inner positive signals in $\boldsymbol{q}_1$ ring (note that the $d$ orbital is defined in 1-Fe unit cell, see Figure S8 for the orbital character of FSs), inter-orbital scatterings of $d_{xz}\leftrightarrow d_{yz}$ normally are suppressed in FeBSs and have negligible effect on $\boldsymbol{q}_2$ ring[16,20]. For the other, the sign-preserving scatterings associated with the folded bands scarcely contribute to the scattering intensity either, due to the weak spectral weight. Thus, it is reasonable for the inner positive-signal arcs, which mainly consist of the inter-orbital scatterings of $d_{xz}\leftrightarrow d_{yz}$ and the scatterings associated with the folded bands, to be missing in PR-$\boldsymbol{q}_2$ scattering pattern (see Supporting Information IV.A for a thorough discussion of $\boldsymbol{q}_2$ ring).